\documentclass[12pt,article]{aastex}
\usepackage{emulateapj5}
\newcommand{\kms}{km\,s$^{-1}$}

\shorttitle{COLA I- radio observations}
\shortauthors{Corbett et al.}
\begin{document}

\title{First Results from the COLA Project- the 
Radio-FIR Correlation and Compact Radio Cores in Southern COLA Galaxies}

\author{E. A. Corbett\altaffilmark{1}}
\affil{Anglo-Australian Observatory, PO Box 296, Epping 1710, NSW, Australia}
\author{R.P. Norris}
\affil{Australia Telescope National Facility, CSIRO, PO BOX 176, Epping NSW, Australia}
\author{C.A. Heisler\altaffilmark{2} and M.A. Dopita}
\affil{Research School of Astronomy and Astrophysics, Australian National University, Private Bag, Weston Creek PO, ACT 2611, Australia }
\author{P. Appleton and C. Struck}
\affil{Erwin W. Fick Observatory and Department of Physics and Astronomy, Iowa State University, Ames, IA 50011 }
\author{T. Murphy\altaffilmark{3}}
\affil{School of Physics, University of Sydney, Sydney, NSW 2006, Australia}
\author{A. Marston}
\affil{Infrared Processing and Analysis Centre, California Institute of Technology, Pasadena, CA 91125}
\altaffiltext{1}{RSAA, Australian National University, Private Bag, Weston Creek PO, ACT 2611, Australia }
\altaffiltext{2}{Deceased}
\altaffiltext{3} {New address: Insitute for Astronomy, Royal Observatory,  Blackford Hill, Edinburgh, UK}

\begin{abstract} 

We present the first results from the COLA (Compact Objects in Low-power AGN) project which aims to determine the relationship between one facet of AGN activity, the compact radio core, with star formation in the circumnuclear region of the host galaxy. This will be accomplished by the comparison of the multi-wavelength properties of a sample of AGN with compact radio cores, to those of a sample of AGN without compact cores and a matched sample of galaxies without AGN.

In this paper we discuss the selection criteria for our galaxy samples and present the initial radio observations of the 107 Southern ($\delta <0\degr$) galaxies in our sample. Low-resolution ATCA observations at 4.8, 2.5 and 1.4  GHz and high resolution, single baseline snapshots at 2.3 GHz with the Australian LBA are presented.  We find that for the majority of the galaxies in our sample, the radio luminosity is correlated with the FIR luminosity. However, a small number of galaxies exhibit a significant radio excess causing them to depart from the FIR-radio correlation. 

Compact radio cores are detected at fluxes $>1.5$mJy in 9 of the 105 galaxies 
observed with the LBA. The majority (8/9) of these galaxies exhibit a significant radio excess and 50\% (7/14) of the galaxies which lie above the radio-FIR correlation by more than $1\sigma$ have compact radio cores. The emission from the compact cores is too weak to account for this radio excess, implying that there are radio structures associated with the compact cores which extend further than the 0.05 arcsec resolution (corresponding to a linear scale 11-22pc) of the LBA. There is no evidence that the radio luminosity of the compact cores is correlated with the FIR galaxy luminosity, indicating that the core contributes little to the overall FIR emission of the galaxy.

The galaxies with compact cores tend to be classified optically as AGN, with two thirds (6/9) exhibiting Seyfert-like optical emission line ratios, and the remaining galaxies classified either as composite objects (2/9) or starburst (1/9). The galaxies classified optically as AGN also exhibit the largest radio excesses and we therefore conclude that a large radio excess on the radio-FIR correlation is a strong indication of an AGN with a compact radio core.   
\end{abstract} 

\begin{keywords} 
\\
Galaxies:active --- galaxies:starburst ---- infrared: galaxies ----- radio continuum: galaxies 
\end{keywords}
\section{INTRODUCTION}

Recent observations of nearby active galactic nuclei (AGN) have revealed that 
some, if not most, AGN host galaxies exhibit circumnuclear starburst activity.  
Examples of this are the circumnuclear ring of star formation found in the 
Seyfert 1 galaxy NGC 7469 (Genzel et al. 1995), the stellar core (Thatte et al. 
1997) and extended ($\approx$ 3kpc from core) star forming region (Telesco \& 
Decher 1988) observed in NGC 1068, and the compact starburst found in the 
extremely luminous Seyfert 2 galaxy, Mrk 477 (Heckman et al. 1997). These 
starburst regions can account for more than 50\% of the galaxy's bolometric 
luminosity (Genzel et al. 1995; Heckman et al. 1997). In addition to starburst 
activity, AGN also exhibit other circumnuclear phenomena  such as  inner rings 
(Baum et al. 1995; Colina et al. 1997), bars (Ho, Fillipenko \& Sargent 1997; 
Shlosman, Peletier \& Knapen 2000) or spirals (e.g. Colina et al. 1997; Colina \& Arribas 1999; Regan \& Mulchaey 1999; Martini \& Pogge 1999). These structures may well be related to the mechanism by which the AGN is fuelled (Ho, Fillipenko \& Sargent 1997;  Maiolino et al. 1997; Shlosman, Begelman \& Frank 1990; Norman \& Scoville 1988).

What is not clear, however, is whether there is a causal connection between the 
circumnuclear activity and the presence of an AGN. It is possible that the 
starburst activity precedes the ``switching on" of the AGN and one might imagine
the scenario of a galaxy interaction causing infall, leading to starburst 
activity which in turn leads to further infall and finally to the formation of 
an AGN (Sanders et al. 1988). Certainly Boyle \& Terlevich (1998) found a 
striking similarity between the evolution of the QSO luminosity density and the 
galaxy star formation rate, indicating that star formation plays an important 
role in QSOs. In fact Terlevich et al. (1992) argue that parsec scale starbursts combined with compact supernova remnants can reproduce the observed properties 
of an AGN, eliminating the need for a black hole completely. It is also possible that the star formation activity is actually triggered by the disruption of gas 
caused by the presence of the AGN (e.g. van Bruegel \& Dey 1993; van Breugal et 
al. 1985) or indeed is unrelated to the presence of the AGN, with mergers 
creating circumnuclear star formation but the AGN activity maintained by  structures which were formed on small scales early on in the galaxy's life cycle.

The primary goal of the COLA project (Compact Objects in Low-power AGN) is to 
determine the relationship between one characteristic of an AGN, the presence of a compact radio core,  with star formation in the circumnuclear region of the host galaxy. This will be accomplished by comparison of the multi-wavelength properties of a sample of AGN with compact radio cores, to comparison samples, consisting AGN without compact cores and a matched sample of galaxies without AGN.

In particular, the circumnuclear regions of the three galaxy groups will be 
examined for evidence of fueling (or fueling suppression) and the relative 
prevalence of nuclear structures such as bars or rings. The internal kinematic 
properties of their gas and the spatial distribution of emission mechanisms such as photoionisation and shock-excitation, e.g. ionisation cones, will also be 
investigated using emission-line imaging and optical spectroscopy. Finally we 
will aim to establish what differences, if any, exist between the AGN with 
compact radio cores and those without, and whether their radio structure at pc-
and kpc- scales is aligned or associated with any optical structures.

In order to avoid biases in our sample we have selected a large initial sample of galaxies from the IRAS Point Source Catalogue (Joint IRAS Science 
Working Group 1988) and determined their properties over a range of wavelengths. It is from this initial sample that the AGN and non-AGN comparison sub-samples will be selected for further study, with the non-AGN galaxies matched to the AGN galaxies using global, orientation independent measures. The selection criteria used for the individual samples and sub-samples will be discussed in detail in Section 2, together with the possible sources of bias.

The initial sample of galaxies has been divided into the Northern COLA sample 
($\delta>0^{o}$) and the Southern COLA sample ($\delta<0\degr$). In this paper 
we present low resolution radio observations and high resolution long baseline 
array (LBA) snapshots of the Southern COLA sample. The low resolution 
observations measure the total radio luminosity of the galaxies whereas the LBA 
snapshots are used to determine which of the galaxies have compact radio cores 
($<$0.08-0.05 arcsec in diameter). We discuss the radio and FIR properties of 
Southern COLA sample as a whole and examine the characteristics of the galaxies 
in which we detect compact radio cores. Optical spectroscopic observations and 
CO spectral line measurements have also been obtained for the Southern COLA 
sample and these will be discussed elsewhere (Corbett et al., in preparation, and Norris et al., in preparation).  Throughout this paper we have assumed $H_{0}$=75 km s$^{-1}$ Mpc$^{-1}$ and q$_{0}$=0.5.

\section{SAMPLE SELECTION AND BIAS}

Previous investigations into the relationship between AGN and the host galaxy 
have been biased by the difficulties associated with defining the AGN sample and the matched non-AGN sample. In fact most AGN samples are biased to some degree, 
having been based on criteria such as UV excess, far-infrared spectral index, 
etc. For example, samples of Seyferts are sometimes selected by their far-
infrared spectral indices (eg. de Grijp et al 1985, Heisler 1991). However, the 
dusty tori that are postulated to exist in Seyferts may have significant opacity even in the mid-infrared (e.g. NGC4945; Brock et al 1988), so that only longer wavelength radiation ($>$ 50 $\mu$m) can be assured of escaping directly. Theoretical models indicate that an increase in the inclination of a dust torus would cause a steepening of far-infrared spectral indices (eg. Pier \& Krolik 1993, Efstathiou \& Rowan-Robinson 1995 ) and this is consistent with observational data (Heisler et al. 1998, Heisler, Lumsden \& Bailey 1997). These Seyfert samples would therefore be biased towards objects orientated at a particular range of inclinations. Similarly, galaxies are often classified as AGN or starburst (HII) based on their emission line ratios, following the empirical system developed by Veilleux \& Osterbrock (1987). However, galaxies often contain both AGN and starburst regions, thus straddling the empirical distinction between starburst and AGN (e.g., Hill et al. 1999; Kewley et al. 2000a). Furthermore, if the nuclear regions are obscured by dust, the starburst component may dominate the observed optical spectrum, masking the AGN. The investigations by Hill et al. (1999) have also  shown that low metallicity AGN have the potential to be misclassified as starbursts. 

Finally, investigations using samples of galaxies can also be affected by 
Malmquist and other distance-dependent biases. For example, the de Grijp et al. 
(1985) sample has a bias towards warmer objects, and because most of them 
are close to the IRAS detection limit, the Seyfert 1 galaxies selected tend to 
be slightly more distant and more luminous than the Seyfert 2 galaxies. 

It is therefore vital that we consider and eliminate as many of these biases as 
possible from the COLA sample. To avoid distance-dependent biases, we have 
selected a large initial sample of galaxies from the IRAS catalogue within a 
well-defined shell of distance (as given by their heliocentric
velocity) and with a far-infrared (FIR) flux limit set sufficiently high for 
them to be detected easily at optical and radio wavelengths.  We have then 
observed these galaxies over a range of wavelengths (a) to determine which 
galaxies are AGN and (b) to measure orientation independent properties of the 
galaxies such as the total luminosity at radio and I-Band wavelengths and the 
total integrated CO luminosity. We will then select the AGN and non-AGN comparison sub-samples for further study, matching the non-AGN galaxies to the AGN galaxies using their orientation independent measures. 

For this study to be successful, we must establish our AGN selection criteria 
carefully.  As discussed previously, AGN seem to possess a circumnuclear  
dusty disk or torus, which can obscur the nucleus along certain lines of sight. Thus, defining AGN on the basis of their IR-UV emission can 
easily introduce an orientation-dependent bias. Additionally, the optical spectral signature of an AGN can be dominated by circumnuclear star formation.
These problems are avoided by observing at radio frequencies. The detection of a high brightness temperature (T$>$10$^5$ K) compact radio core is a strong 
indication of AGN activity (Norris et al. 1990), and since the effect of dust is negligible at radio wavelengths, it is independent of orientation.  AGN with such compact radio cores exhibit a radio excess on the radio-FIR correlation (Roy et al. 1998). Since the FIR emission is dominated by star-formation activity in the galaxy disk, a radio excess indicates that the AGN rather than 
the starburst dominates, at least at radio wavelengths, and should be sufficiently powerful to influence the circumnuclear regions of the host galaxy. We have therefore chosen to use the presence of a compact radio core as our principle diagnostic of nuclear activity. 

\subsection{Sample Selection Criteria}

The COLA sample consists of all the objects in the IRAS catalogue with a 
heliocentric velocity, as given by Strauss et al. (1992), between 3500 and 7000 km/s ($0.0117 < z < 0.0234$) and a flux at 60$\mu$m, S$_{60}$, $>$ 4 Jy. This cut-off limit was chosen as the well known FIR-radio correlation for galaxies indicates that a galaxy in the luminosity and redshift range of this sample should have  a 13-cm radio flux in excess of 10 mJy (Roy et al. 1998) and will therefore be detected easily at radio wavelengths with a total on-source integration time of 5-6 minutes. We further imposed the criterion that the galactic longitude $\vert b \vert > 10^\circ$ so that the far-infrared data are not confused by Galactic cirrus. The resulting sample contains 217 objects of which 107 have $\delta <$ 0$^{\circ}$, i.e. are Southern Hemisphere objects. 

Choosing an appropriate flux limit  for the detection of a compact radio core in an AGN is not quite so straight forward. An AGN may contribute any proportion of the total radio flux of a galaxy; for example, the contribution of a weak AGN like that at the centre of our Galaxy would be undetectable in our sample whereas a radio-loud quasar would dominate the emission from the host galaxy. We cannot therefore distinguish cleanly between AGN and non-AGN galaxies, but must instead choose a threshold level of AGN activity above which the AGN is sufficiently dominant to influence with the host galaxy. The challenge is to choose this threshold in an unbiased way.

For the Southern COLA sample, we have elected to set the detection limit of a compact core to be 1.5mJy at 13cm with 0.05 arcsec resolution (i.e. the ATCA-Tidbinbilla baseline on the Australian LBA). At this flux level, our FIR selection criterion together with the radio-FIR correlation ensures that we will detect any radio core which emits at least 15\% of the radio flux expected from star formation activity. This detection threshold is well-matched to optical spectroscopic classification, since Kewley et al. (2000b) have shown the spectrum of a galaxy containing an AGN with 15\% of the luminosity of the starburst activity in the galaxy would be spectroscopically classified as an AGN. We cannot, of course, rule out the existence of a very low-power AGN below our sensitivity limit. However, such objects are unlikely to dominate either our control sample or the energetics of the host galaxy in which they reside.   The LBA observations are made at 13cm wavelengths (2.3  GHz) because, although these objects are generally brighter at longer wavelengths with a mean spectral index 
of $\sim$ 0.7, free-free absorption (or synchrotron self-absorption) can reduce the observed flux at wavelengths from 20cm above.

\subsection{Potential Sources of Bias}

\subsubsection{The 60$\mu$m Flux Limit} 
 
Imposing the FIR flux cut-off using IRAS data results in the selected galaxies having FIR luminosities $>10^{10.5}$ L$_{\sun}$, significantly larger than for ``normal'' galaxies (e.g. the Shapely-Ames sample). Since there is a correlation between FIR luminosity and the 60/100$\micron$ color it is to be expected that that galaxies in the COLA sample will be warmer than galaxies selected without an IR cut-off.  This is indeed the case as the distribution of IRAS 60/100$\micron$ color temperatures for the COLA sample shows that the COLA galaxies are, on average warmer than samples of ``normal'' galaxies (Fig. 1). 

The high FIR luminosity of the COLA galaxies also results in an
enhanced fraction of interacting galaxies in our sample. For example, from V-band images obtained at the E. W. Fick Observatory, we
estimate that out of a total of 111 Northern COLA galaxies, 44 (40\%) are obviously involved in either tidal interaction or a major merger, and a further 23 (20\%) are in apparently non-interacting, but often widely separated pairs (Appleton et al., in preparation).

Finally, the IRAS telescope was only capable of resolving emission from galaxies if they were separated by more than 4 arcminutes at 100$\micron$. Hence, in paired systems both galaxies are often within the IRAS beam. It is unknown whether either of the galaxies would have met the selection criterion if isolated and it is therefore impossible to determine whether this will introduce significant bias into our sample.  

\subsubsection{Non-AGN cores}

Observations by Kewley et al. (2000a) of a sample of 61 luminous infrared galaxies have revealed compact radio cores ($<0.1$ arcsec in diameter) in 37\% of galaxies classified as starburst from optical spectroscopic observations and 80\% of the galaxies classified optically as AGN. The radio luminosity distribution of the compact cores was found to be bimodal with the compact cores in the starburst galaxies tending to be less luminous ($< 10^{4}L_{\sun}$) than those detected in the AGN. Although the radio cores in these starburst galaxies could be obscured AGN, Kewley et al. concluded that they were more likely to be complexes of extremely luminous supernovae such as those seen in 
Arp 220 (Lonsdale, Smith \& Lonsdale 1993; Smith, Lonsdale \& Lonsdale 1998).  

We therefore need to avoid contaminating the COLA sample with such objects and detect only bona fide AGN. Detection of a compact core depends, of course, on its intrinsic luminosity and redshift. With a detection limit of 1mJy and a resolution of 0.1arcsec, the galaxies optically classified as starburst with compact cores make up less than 15\% (1/8) of the detected objects at a redshift $>$ 0.01 in the Kewley et al. sample. 

Given that the SNR candidates tend to have compact cores $< 10^{4}$ L$\sun$, corresponding to 1.5mJy at the upper redshift limit of our sample (7000 \kms), raising the detection limit to 1.5mJy for the COLA sample should reduce the number of these objects in our sample. Additionally, the increased resolution of the ATCA-Tidbinbilla LBA (0.05 arcsec, equating to 15pc at the lower limit of the redshift range of our sample) may resolve out some of the members of  supernovae complexes, resulting in fewer SNR candidates being detected. We therefore believe that the LBA detections for the COLA galaxies will not be significantly contaminated by these putative complexes of supernovae.
 
\subsubsection{Relativistic Beaming}

Using compact cores to define AGN activity does have some potential to introduce bias into the COLA sample. If, as is believed to be the case for the majority of objects (Ulvestad et al. 1999, Mundell et al. 2000), the compact radio cores 
observed in Seyferts are the base of a radio jet, relativistic beaming may 
introduce an orientation bias. The relativistic velocity of the particles in the jet will direct the majority of the emission along the jet axis and we are therefore more likely to detect Seyferts with radio jets aligned close to the line of sight than those in which the radio jet is perpendicular to the 
observer. In the context of AGN unification this would result in more cores 
being detected in Seyfert 1s than Seyfert 2s. Nagar et al. (2000) detected 
compact ($<$0.2 arcsec) cores at 15  GHz in 57\% of their sample of Seyferts, 
LINERS and ``transition objects'' (i.e. galaxies whose optical spectral line ratios lie on the AGN/starburst  partition) and 83\% of the sources detected had flat spectra, usually indicative of synchrotron self-absorption and often taken to mean that the jet axis is closely aligned to the line-of-sight. However, Roy et al. (2001) have shown that these flat-spectrum cores are more likely to be caused by free-free absorption than by synchrotron self-absorption.
Furthermore, Roy \& Norris (1994) have found that in their survey of over 200 Seyfert galaxies, compact cores appeared to be more common in Sy 2s than Sy 1s. In fact there appears to be growing evidence that those radio jets observed in Seyferts are not relativistic (Ulvestad et al. 1999), removing the possibility that Doppler boosting of the core emission will introduce an orientation bias to our sample. 

It therefore seems unlikely that relativistic beaming will cause a bias towards 
objects with jets close to the line of sight, but any evidence for a higher 
number of Seyfert 1s than Seyfert 2s in our survey, or a higher detection rate 
for compact cores in flat spectrum sources, would be a cause for concern.   

\subsubsection{Variability}
It is possible that the compact cores in our sources may be variable. In general, radio quiet galaxies are not highly variable at radio wavelengths, as is demonstrated by the repeated observations of well-known sources such as NGC1068, and the overall consistency of repeated observations similar to those presented here by Roy et al. (1994), Kewley et al. (2000), and Norris et al. (2001). However, some examples of variability are known. For example, Wrobel (2000) found that the 8.4  GHz flux density of the central nucleus of the Seyfert 1 galaxy, NGC 5548, varied by about $\sim$50\% over a period of 4.1 yr and 33\% over a period of 41days at  500mas and 1250mas resolution. This variability was reduced to 19\% at 4.9  GHz. NGC 5548 is well known as a highly variable Seyfert and it is unlikely that any of the sources in our survey will display such extreme behaviour. 

Falcke et al. (2000a,b) and Nagar et al. 2000 have conducted a study of the radio properties of two different sets of AGN. The first group consisted of 30 objects selected as representative of the radio-quiet, radio-loud and radio intermediate classes of AGN. With a few exceptions the great majority of these objects have a varibility of less than 10\% (when normalised to take into account the measurement errors). Their second group of objects (Falcke et al. 2000b) consisted of a sample of low luminosity, nearby AGN . They found that the incidence of variability in these objects was much higher than for the first sample, although again the majority of the objects ($\sim$ 55\%) exhibited variability of 25\% or less. None of these objects exhibited variability of more than 70\%.

Variability of 20 - 50\% in the compact cores of our galaxies is unlikely to make a statistical difference to our sample as a whole. The worst possible scenario would be that an AGN would happen to have a core flux below our detection limit when we observed it, but this would only occur if the AGN flux was usually close to our detection limit, i.e 50\% variability in a 1.5 mJy source might cause us to reject it should we observe it in a weak state whereas a 4mJy compact source which exhibits the same level of variability would always be detected. We therefore believe that radio variability will have little effect on the outcome of this study but we intend to return to some sources to check for significant variations.  

\subsubsection{Free-free absorption}

The advantage of using radio observations to probe the nuclei of starburst and 
active galaxies is that radio waves are unaffected by the many magnitudes of 
extinction due to dust in the cores of these galaxies. However, radio waves are 
subject to free-free absorption by ionised gas along the line of sight. 
Generally, the densities of ionised gas are not sufficient to cause appreciable 
free-free absorption at centimeter wavelengths, but Condon et al. (1991) 
reported a deficit in radio emission of some starburst galaxies at 1.49  GHz, 
which he interpreted in terms of free-free absorption. Free-free absorption has 
since been seen directly in VLBA images of Seyferts by Roy et al. (1999; 2001). 
However, it is important to note that in no case did free-free absorption render a Seyfert galaxy invisible to VLBI baselines. Instead, the free-free absorption 
covers only a part of the source in the images produced by Roy et al., and in 
all of those cases the source would still be detectable by the observations 
described here, although at a slightly lower flux density than would be observed in the absence of free-free absorption. 

\section{OBSERVATIONS}

\subsection{ATCA observations}
The Australia Telescope Compact Array in the 750 configuration was used to 
obtain low resolution images of 107 galaxies in our sample at 4.8  GHz, 2.5  
GHz and 1.4  GHz (6-, 13- and 21-cm). 
Data from Antenna 6 (the 6 km telescope) were excluded 
and hence the largest baseline was 643m. This compact configuration was chosen so that the extended emission from the galaxies would be unresolved and thus the total integrated flux of the galaxies could be measured easily. The 4.8 GHz observations were conducted on 1998 June 23 using 2 IFs with a bandwidth of 132 MHz and central frequencies of 4.8 GHz and 4.93 GHz respectively. The 2.5  GHz and 1.4 GHz observations were obtained simultaneously on 1998 July 8-9 \& 27-31 with a bandwidth of 132 MHz at each frequency. Additional 2.46 GHz (132MHz bandwidth) observations of 53 sources were obtained in 1998 September 19-20 during our LBA observing program and were combined with the 1998 July data to improve the UV-coverage. 
 
The ATCA configuration gave a nominal beam diameter of 10 arcsec but, due to the difficulties involved in obtaining images of more than one hundred galaxies over a wide range of declinations, the beam was often highly ellipitical. Thus the major axis of the beam was often between 100-150 arcsecs (40-60Kpc at the redshift range of our sample) at 4.8GHz.

The galaxies were typically observed for 120s per scan at 4.8 GHz and 80s at 
2.5 GHz and 1.4 GHz. A total of 6 scans per source were taken at different 
hour angles. The primary calibrator was PKS 1934-638 and secondary calibrators were observed regularly at approximately 20 minute intervals. The data were reduced and calibrated according to standard AIPS procedures and the central region containing the target was cleaned and mapped using the CLEAN fit algorithm, with no self calibration performed. The mean rms sensitivity was typically 0.7 mJy at 4.8 GHz, 0.6mJy at 2.5  GHz and significantly larger at 1.4 GHz cm ($\sim$ 1mJy). 

At the resolution of the observations, the vast majority of the galaxies were point sources and hence their flux densities were estimated by fitting a two dimensional Gaussian to the cleaned images and measuring the peak and integrated flux density. Sources were detected if the peak and/or integrated flux exceeded 5 times the RMS level of the map. 
  
\subsection{Australian LBA Observations}

Snapshot observations of the southern COLA galaxies were obtained with the 
Australian Long Baseline Array (LBA) to search for compact cores. The 
observations split into two observing runs with similar setups and separated by 
nearly two years. On  1998 September 19 \& 20, 53 galaxies were observed at 2.3  GHz with an effective bandwidth of 16 MHz using the ATCA - Tidbinbilla 
telescopes. These telescopes form a 566km baseline providing a 0.05 arcsec 
fringe-spacing. Each target source was observed for between 20$\sim$24 minutes,  and only the right hand circular polarization state was obtained from the receivers. Phase calibrators, selected from Duncan et al. (1993) or the LBA calibrator source list, were observed for approximately 
10 minutes every 4-8 hours during the run. These were all objects known to have compact cores brighter $<$1Jy and believed to have little extended flux.  
In 2000 July LBA observations of the remaining COLA galaxies were obtained, again at 2.3 GHz with a 16 MHz bandwidth and this time both left and right polarisation data were collected. Due to scheduling constraints only 42 galaxies were observed with the ATCA-Tidbinbilla baseline with the remaining 12 galaxies observed using the ATCA and Parkes telescopes, providing a baseline of 321km (0.08 arcsec resolution). There was an overlap period of 5.5hrs when all three telescopes were used, but for consistency only the ATCA-Tidbinbilla baseline data, when available, are presented.  Four galaxies observed during the 1998 LBA run were re-observed in order to provide internal calibration of the two samples. For the ATCA-Tidbinbilla observations, the total on source integration time was 25 minutes and phase calibrators were observed for 10 minutes at intervals of 3-4 hours. For the ATCA-Parkes observations the total on source integration time was approximately 45 minutes with phase calibrators observed for 10 minutes at intervals of 4-5 hours. 

The observations were correlated using the ATNF S2 correlator and reduced in the standard fashion. To remove residual delays and phase errors, the fringes were 
searched in delay space using the AIPS procedure ``FRING'' (Schwab \& Cotton 
1983). Fringe detections with a S/N ratio greater than 5 were then used to 
calibrate the delays via the AIPS procedure ``CLCAL'', interpolating between 
solutions with a low order polynomial. At this stage, the data were corrected for flux variations over the bandpass, but no absoulte flux calibration was applied.  

The frequency channels were then averaged spectrally and split into the 
individual sources. To determine whether a compact core had been detected, an AIPS procedure called {\it ``FRPLOT''} was used to plot the fringe rates of the data with a detection indicated by a large spike in flux close to the phase center. Each polarization and, in the case of those sources observed with all three telescopes, each baseline was examined separately. The total flux of the detected compact cores was measured by averaging the visibility data so that bad data caused, e.g. by the temporary loss of one antenna or delays in driving the telescope, could be detected by eye and flagged. 

The majority of fringe rate detections for both the target sources and 
calibrators were found to be within 2mHz of the phase center, but a few sources 
exhibit detection spikes up to 17mHz from the zero-point. The main cause of this was a shift between the actual source position and the position for which the 
data had been correlated. For the 1998 run, the positions derived from IRAS 
measurements were used to set the phase centers and due to the uncertainty in 
these measurements it is not surprising that there was some discrepancy between 
the radio and the IR positions. Additionally it is also possible that the 
polynomial fit to the solutions from the fringe fitting was not optimal and 
caused a slight shift. For the 2000 observations, the source position measured 
from the 4.8 GHz radio data was used as the phase center for the correlations 
and the largest shift in the fringe rate detections was 5mHz. We believe this 
shift to be due to errors in the delay calibration rather than source position 
as it is only seen in two sources which were observed one after the other. Any large ($>2.5$mHz) shifts in the fringe rate spectrum were corrected  
before measuring the core flux from the visibilities by 
shifting the phase center in RA and Dec. The largest shift necessary was 23 arcsec in RA (applied to data obtained in 1998) and shifts were only applied to 4 sources. 

It is not possible to obtain an absolute flux calibration from the LBA observations alone as the calibrators with compact cores, by their nature, are variable and the standard primary calibrator for Southern Hemisphere observations, PKS 1934-638, is resolved on the LBA. We therefore used the data collected on the ATCA during the LBA observations to provide an absolute flux calibration in the following manner. The ATCA data was correlated and the data reduced following standard AIPS and ``MYRIAD'' procedures. PKS 1934-368 was used to calibrate the flux of the compact calibrators. Since the calibrators were only observed once during the 24+ hour observing runs, we must assume that the antennae gains varied only by a small amount during the run. The ATCA fluxes of the compact calibrators were then compared with the fluxes measured on the LBA baselines (ATCA-Tidbinbilla in 1998 and Parkes-ATCA, ATCA-Tidbinbilla and Parkes-Tidbinbilla in 2000). 

Although the compact calibrators are known to possess little extended flux, it is possible that they are surrounded by a low surface brightness halo (or ''fuzz'') of emission which could be detected by the ATCA in such a compact configuration but is resolved out with the longer baselines of the LBA. It is therefore to be expected that slightly larger fluxes for the calibrators would be measured on the ATCA than the LBA. From comparison of the ATCA and LBA observations, it did indeed appear that several of the calibrator sources had some low flux density extended emission and they could therefore not be used to calibrate the data. The remaining calibrators, which we believed to be unresolved at all baselines, were used to flux calibrate the data. For the 2000 July observations, the ATCA-Tidbinbilla data were scaled by a factor of 1.06 and the Parkes-ATCA data by 1.08. The 1998 September data, which used only the ATCA-Tidbinbilla baseline, were scaled by 1.13. 

The rms variability in the LBA visibilities of the calibrator sources was less than 0.5\% of the mean flux for the unresolved sources. 

\subsection{Far Infra-red Data}

The far infra-red (FIR) flux densities at 10, 25, 60 and 100 $\mu$m of the 
sample were taken from the IRAS Point Source Catalogue, Version 2 (Joint IRAS Science 
Working Group 1988). The flux densities were converted to a ''Far-infrared'' 
flux, $S_{FIR}$, following Helou, Soifer \& Rowan-Robinson (1985), by the 
relationship
\begin{equation}
S_{FIR}=1.26 \times 10^{-14}[2.58 S_{60} + S_{100}]/3.75\times 10^{12}
\end{equation}
where $S_{60}$ and $S_{100}$ are the 60$\mu$m and 100$\mu$m flux densities 
respectively. Errors in these fluxes are typically 6-10\% (IRAS Point Source Catalogue).

The fluxes at 10 and 25 $\mu$m were excluded as there is some evidence (as discussed in the preceeding section) that they can introduce an orientation bias into the data.

\section{RESULTS}
\subsection{ATCA Observations}
The radio flux measured from the ATCA images is shown in Table 1. The median difference between the integrated flux and the peak flux of the Gaussian fit is 0.04mJy (with a standard deviation of 0.3mJy) for our sample as a whole, as would be expected for a sample of generally point-like unresolved sources. In some cases the integrated flux was between 20-70\%  higher than the peak flux. These objects were found to be resolved at 4.8 GHz 
and have been classified as extended in Table 1. For around half the objects the integrated flux was actually slightly smaller (median difference 0.9mJy or 8\% 
of the peak flux) than the peak flux. These sources were unresolved at 4.8  GHz 
and we believe that the disparity between the integrated and peak flux reflects 
the error in the Gaussian fit to the data. For these objects the peak flux was 
used in the subsequent analysis instead of the integrated flux. We estimate conservatively that the error in the fluxes derived from our Gaussian fits to the data is 10\% at 4.8  GHz and 2.5 GHz rising to 15\% at 1.4  GHz.   

Fluxes were not measured for all 107 objects at 4.8, 2.5 and 1.4 GHz due to the difficulty of scheduling observations over a sufficient range of hour angles number to map such a large number of sources. This was a particular problem at 1.4  GHz since additional observations were not obtained during the 
September LBA run. Three sources were not detected (at the $5\sigma$ level) at 4.8  GHz, ten at 2.5  GHz and nineteen at 1.4  GHz. The large number of non-detections at 1.4  GHz is at least partially due to the higher noise in the 1.4  GHz maps caused by the shorter integration times and smaller UV-coverage of the 1.4  GHz observations relative to those at 4.8  GHz. Finally, because of the increased beam sizes at 2.5 and 1.4  GHz, we encountered some difficulty in separating the nearby sources in from the targets and a further three sources at 2.5 and four sources at 1.4  GHz were excluded from the subsequent analysis. 
 
In total, flux density measurements were obtained for 104 objects at 4.8 GHz, 94 at 2.5 GHz and 84 at 1.4 GHz.

\subsection{LBA Observations}

The ATCA - Tidbinbilla LBA is sensitive only to structures smaller than 0.05 
arcsec, corresponding to $15 - 25$ pc over the redshift range of our sample, 
and brightness temperatures of $> 10^{5}$K. It will therefore not detect 
emission from the kpc-scale starburst regions, which have typical brightness 
temperatures of $10^{4}$K. The ATCA-Parkes baseline is somewhat shorter and is 
sensitive to structures smaller than 0.08 arcsec, again with a brightness 
temperature $>10^{5}K$ (due to the decreased sensitivity). The single baseline, snapshot technique of detecting compact cores does not allow accurate measurement of the core positions.

Compact radio cores were detected in 9 objects with the ATCA-
Tidbinbilla baseline (Table 1) and upper limits obtained for the remaining 96 
sources. Eight of the detections were at fluxes $>$2mJy and, when data for both 
polarisations was available, were seen at the same fringe rate in both 
polarisations. A tentative detection was obtained from the July 
2000 data for IRAS 12329-3958. Identical, but low amplitude, features are 
seen in both polarisations of the IRAS 12329-3958 data. The amplitude of the 
feature is similar to the surrounding noise but the fact that the detection 
(made up of 6 data points in total) is exactly the same in both polarisations 
seems unlikely to be a coincidence. In the subsequent analysis, statistical 
tests are performed twice, once for all nine of the galaxies in which compact cores were detected and once with IRAS 12329-3958 excluded. The main results 
and conclusions are unchanged. 

Four sources, IRAS 19543-3804, IRAS 09375-6951, IRAS 12329-2958 and IRAS 04118-3207, were observed in both LBA runs. Compact cores were detected in IRAS 19543-3804 and IRAS 09375-6951 during both LBA runs and the flux measured on the ATCA - Parkes baseline in 2000 was the same or higher than that obtained in 1998 with the ATCA-Tidbinbilla baseline. As discussed above a tentative detection was made for IRAS 12329-3958 on the ATCA-Tidbinbilla baseline in 2000 by comparing the fringes of two polarisations measured. Since only one polarisation was measured in 1998, the compact core was not detected during that run. No compact core was detected in IRAS 04118-3207 in either observing run.   
 
The average upper limit for non-detections was 1.7 mJy for the ATCA-Tidbinbilla baseline (both in 1998 and 2000) and 2.5 mJy  for the Parkes-ATCA baseline. The rms of the visibilities measured for detected objects was 0.2mJy for each baseline. 
  
\subsection{Radio-FIR Correlation}

Most galaxies exhibit non-thermal radio emission which appears to be correlated 
with their thermal far-infrared emission. This correlation is well-known and 
extends over several orders of magnitude from "normal" spirals  (e.g. Dickey \& 
Salpeter 1984; Helou, Soifer \& Rowan-Robinson 1985; Wunderlich, Klein \& 
Wielebinski 1987; Sopp \& Alexander 1991; Yun, Naveen \& Condon 2001) to  Seyfert galaxies, albeit with a much greater scatter (e.g. Sanders \& Mirabel 1985; Norris, Allen \& Roche 1988; Wilson 1988; Baum et al. 1993; Roy et al. 1998). It is believed to be indicative of star formation activity and  breaks down for those AGN with compact ($20\sim 200$ pc) radio cores, i.e some radio-quiet AGN and all radio-loud AGN (e.g. Impey \& Gregorini 1993 and Roy et al. 1998), as they exhibit a radio excess.

In normal galaxies non-thermal radio emission is generated in the galactic disk, probably from the non-thermal emission associated with supernova remnants, and is generally coincident with optical indicators of star-formation activity. This source of radio emission is also found in Seyferts and is probably responsible for the FIR-radio correlation observed. The compact high-brightness-temperature radio emission also observed in some Seyferts is indicative of nuclear activity and in general is attributed to non-thermal synchrotron radiation from the core and radio jets. It is this nuclear component of radio emission which disturbs the FIR-radio correlation in Seyferts (e.g. Sanders \& Mirabel 1985; Wilson 1988; Baum et al. 1993; Roy et al. 1998, Yun et al. 2001). Baum et al. (1993) have found that subtraction of the inner 1 kpc of radio emission from Seyferts returned them to the FIR-radio correlation. However, significant nuclear FIR emission has been observed in some Seyferts (e.g. Edelson 1987; Spinoglio \& Malkan 1989; Clavel, Wamsteker \& Glass 1989) and it is therefore not clear whether Seyferts should 
return to the normal-FIR correlation without the removal of this FIR nuclear 
emission as well. Certainly, Roy et al. 1998 found that the subtraction of the 
central 20$\sim$200 pc of nuclear radio emission from those objects with radio 
cores did not return them to the FIR-radio correlation indicating that these 
objects exhibit larger scale radio structures, perhaps analogous to the radio 
jets seen in radio-loud AGN and the linear structures observed in some
Seyfert nuclei.    

The strong linear correlation between the total radio luminosity and the FIR emission of the COLA galaxies can clearly be seen in Fig. 2. This radio-FIR correlation is formally significant with a probability, P$<$0.1\%, from both the Rank Spearman and Kendall $\tau$ test of it being due to chance (obtained from the 4.8GHz data, which is the most complete data set). Since similar results are obtained from the statistical tests performed on the flux-flux data, it is not an artifact caused by presenting the data on a luminosity-luminosity scatter diagram.   

Following Helou et al. (1985), we obtain the quanitity $q=log(S_{FIR}/S_{radio})$ for our sample (Table 2) and $\langle q\rangle$, the median 
FIR to radio flux ratio (Table 3). The wavelength dependence of $q$ is due to the power law relationship between flux and frequency at radio wavelengths. 

Inter-comparison of different studies are often hampered by the fact that the surveys were performed at different radio frequencies. In the past, the radio fluxes measured at different frequencies have been converted by assuming a power-law relationship, S$\propto \nu^{-\alpha}$, with $\alpha$ taken to be 0.7 (e.g. Roy et al. 1998) or, in  the case of the 60$\mu$m peakers, -0.3 (Heisler et al. 1998). Since we have observations at all three frequencies however, we can do a direct comparison between our results and those in the literature. 

We find that the median $q$ for the COLA galaxy sample at 4.8  GHz, $\langle q_{4.8} \rangle=2.83\pm0.03$ is slightly higher than that found by Wunderlich et al. (1987) for a sample of normal spirals at 4.8 GHz, $\langle q_{4.8} \rangle=2.71\pm$ 0.03. It is also significantly higher than that of the 36 Seyferts ($\langle q_{4.8} \rangle=2.44\pm$ 0.03) in their sample but lower than their sample of 50 HII regions ($\langle q_{4.8} \rangle$=3.12). This 
slight FIR excess is also seen at 2.5 GHz as $\langle q_{2.5} \rangle$ for the COLA sample, ($2.60 \pm 0.03$) is marginally higher than that obtained by Roy et al. (1998) for their sample of Seyfert galaxies without compact radio cores ($\langle q_{2.4}\rangle=2.48\pm$0.1) and at 1.4  GHz, where we measure $\langle q_{1.4} \rangle=2.43\pm0.03$ compared to the $\langle q_{1.49} \rangle=2.34$ measured by Condon, Anderson \& Helou (1991) for their sample of galaxies taken from the Bright Galaxy Sample.  This is comfirmed by recent results from Yun et al. (2001) who obtain a mean $q_{1.4} = 2.34 \pm 0.01$ from 1809 objects with IRAS $S_{60} \ge$2Jy.  

It therefore appears that the COLA sample is either slightly FIR loud or radio 
quiet. Condon et al. (1991) mapped 40 ultraluminous infrared galaxies at 8.4  
GHz with VLA and found that 25/40 of the galaxies were diffuse at 0.25 arcsec 
resolution and obeyed the FIR-radio correlation, with a median $\langle q_{1.49.} \rangle=2.34$, consistent with star-formation galaxies. The remaining 15 were dominated by compact (but usually resolved radio sources) and appear ``radio-quiet" relative to the extended sources, with 8 having a $\langle q_{1.49.} \rangle 2.6$. This radio-quietness was attributed to free-free absorption at 1.49  GHz and when the data were corrected by scaling the 8.45 GHz flux density to 1.49 GHz, assuming a spectral index of $\alpha \sim$ 0.7, the compact sources, with one exception, had the same $q_{1.49}$ as the starburst galaxies. The one exception being Mkn 231 which was the one source unresolved at 0.25 arcsec and, after correction for free-free absorption, stands out as being radio-loud.

Tempting as it is to invoke free-free absorption to explain our somewhat higher 
value $q_{1.4}$ this cannot be the case for the COLA galaxies as the median two point spectral index between 4.8 GHz and 1.4 GHz (Table 2) is $\sim$ 0.74$\pm$0.03 (rather steeper than one would expect for sources affected by free-free absorption and typical for star-forming galaxies) and a similar FIR-excess is seen at 4.8 GHz. 

The relatively large value of q measured for the COLA sample may be due to a selection bias in our sample.  First, as discussed in Section 2.2.1, it is possible for two galaxies to lie within the IRAS beam and a single IR flux recorded. These galaxies would be resolved in the radio, since the beam size is smaller, and thus radio fluxes would be measured for each galaxy. The ratio of the FIR flux of both galaxies to the radio flux of only one of the galaxies would tend to give a q value which is larger than expected. Additionally, imposing a S$_{60}\mu$m cut-off will also bias the COLA sample at the low luminosity end towards galaxies which are slightly FIR-loud. Finally, the COLA galaxies are actively star- forming and may therefore have an intrinsically higher value of q, probably due to an excess of hot dust and gas. There is evidence from other galaxy samples (Norris, private communication) that q increases with the starforming activity of the galaxies but we find no evidence in our sample of a dependence of q on the FIR luminosity (Figure 3). We note however that the Yun et al. (2001) sample has a $S_{60}\mu$m cut off limit of 2Jy and no redshift limit whereas as our sample's limit is 4Jy.   

It is possible, however, that free-free absorption in the nuclear regions could 
account for the six galaxies which have a spectral index between 4.8 GHz and 
1.38 GHz $<$ 0.4.   

\subsection{Galaxies with compact radio cores}
  
In Figure 4 we have divided the COLA galaxies into those in which radio cores were detected in our LBA observations and those in which no significant core emission was detected (typically $<$ 1.7mJy). Also shown is a line representing $\langle q_{4.8} \rangle$ and two lines representing $\langle q_{4.8} \rangle\pm 1\sigma$ respectively.

It is clear that the galaxies with compact cores are ``radio-loud'' relative to the median $q$ of the COLA galaxies and at 4.8 GHz seven of the nine galaxies in which we detect compact cores lie more than 
1$\sigma$ away from $\langle q_{4.8} \rangle$, the two exceptions being IRAS 13135-2801 and IRAS 13097-1531.  In order to obtain a measure of ``radio - loudness", 
$\langle q_{4.8} \rangle$ can be used to "predict" the total radio luminosity at 4.8 GHz which 
would be observed if all the radio emission were due to star formation,  
$L_{P}$. The radio excess, $R$, can then be defined as 
\begin{equation}
R=(L_{O} - L_{P})/L_{P}
\end{equation}
where $L_{O}$ is the observed luminosity of the galaxy. A positive value of R 
reflects an excess of emission and a negative value a deficit. The 4.8 GHz 
radio excess of the objects in which we detect compact cores is shown in Table 
4 and indicates that only one object IRAS1335-2801, lies below the predicted radio luminsoity, while the seven out of the remaining eight galaxies exhibit radio luminosities generally between 1.6 and 35 times that predicted. A similar radio excess is seen at 2.5 (Fig. 5) and 1.38 GHz and the $\langle q \rangle$ for the galaxies with compact cores is significantly lower than that of the galaxies in which compact cores were not detected (Table 3). A number of different statistical tests were used to confirm that the excess of radio emission in the galaxies with LBA cores is formally significant, with a probability of $<$0.1\% (from the Mann-Whitney or 
Wilcoxcon test on the 4.8 GHz data) of it being due to chance. This result is 
unchanged if we exclude the tentative detection in IRAS 12329-3938.  

This radio excess is not entirely accounted for by the presence of the compact 
radio cores. Subtraction of the core luminosity from the total radio luminosity 
(Fig. 5) gives $\langle q_{2.5} \rangle=2.26$ for the galaxies with compact cores implying that radio structures associated with the central nucleus extend further than the 0.05 arcsec (15-25pc at these redshifts) resolution of the ATCA-Tidbinbilla baseline. The most likely candidate for this emission would be small jets and lobes within the central 100 - 1000 pc of the galaxies, and in fact VLA observations of one of these objects, IRAS 13197-1627 have revealed a linear structure extending about 278pc from the core, believed to be a synchrotron jet (Kinney et al. 2000). This confirms the results of Roy et al. (1998) who found that Seyfert galaxies with compact radio cores ($\theta$ $<$ 0.1 arcsec) depart from the normal correlation, being more radio loud than normal spirals with the same FIR luminosity.  They also found that subtraction of the compact radio cores (i.e. the central 20$\sim$200pc) did not return them to correlation. However, Baum et al. (1993) found that subtraction of the inner 1kpc of radio emission from Seyfert galaxies did return them to the correlation, indicating that the structures responsible for the radio excess are confined to the inner kiloparsec of the galaxies.  

There is no significant difference between the 4.8 GHz to 1.4 GHz spectral index of the galaxies with compact cores, median $\alpha$=0.745, and galaxy sample as a whole, median $\alpha$=0.742. This measurement reflects the spectral index of 
the {\it total} radio luminosity of the galaxies, rather than that of the compact core, but given that the majority of the galaxies with compact cores exhibit a substantial radio excess, presumably from structures associated with the compact cores, the spectral index of this additional emission would be reflected in the spectral index of the total radio luminosity.  We therefore believe that the compact cores have not been preferentially detected in objects with a flat radio spectrum, implying that relativistic beaming of the jet is not introducing an orientation bias in our sample.    
  
Only five of the nine galaxies in which we detect compact cores have published optical spectral classifications from BPT emission line diagrams (e.g. Veilleux \& Osterbrock 1987; Veilleux et al. 1997).  However, in the course of the COLA project we have obtained optical spectroscopic observations with the MSSSO 2.3m telescope of the Southern COLA sample and determined the spectroscopic classification of the galaxies using BPT diagrams. The results for the whole sample will be discussed elsewhere (Corbett et al. in preparation) but the classifications of the 9 galaxies in which compact cores are detected are shown in Table 4. Six galaxies with compact cores are Seyfert AGN (1 Sy1 and 5 Sy2), two galaxies lie on or close to the AGN/HII partition and one galaxy is classified as a starburst or HII galaxy. 

Hill et al. (1999, 2001) and Kewley et al. (2000b) describe how the optical emission-line ratios can be used to determine the relative contribution of stars and Seyferts (modelled as radiative shocks with photoionized precursors) to the ionization of the gas. Effectively, the further away from HII/AGN partition line on the BPT diagrams an object is the larger the Seyfert (or shock) contribution to the observed spectrum. The objects classified as Seyferts, by definition, lie further from the HII/AGN partition line than those classified as borderline objects or HII galaxies. 

In our sample, the Seyfert galaxies with compact cores exhibit larger radio excesses (R=0.85 to 34) than the borderline glaxies (R=0.68, 0.27) and the starburst galaxy (R=-0.17). This implies that the radio excess is a measure of the relative contributions of the star formation activity in the host galaxy and AGN to the observed radio continuum, thus a large radio excess (or small q) indicates that the AGN will also dominate the optical emission line spectrum, whereas a small excess (e.g. that of IRAS 09375-6951) indicates that the AGN will contribute little to the observed optical spectrum ($\sim$10 per cent) and the galaxy will lie close to the AGN/HII partition on the BPT diagrams. This relationship will be investigated further when we present optical spectra of the whole galaxy sample.  

Yun et al. (2001) define ``radio AGN'' as those galaxies whose radio luminosity is more than 5 times that predicted from the radio-FIR correlation. For their large sample of objects this corresponds to galaxies with $q_{1.4}\le 1.64$ and they find that only 1.3\% of their sample fullfil this criterion and are ``radio AGN''. Using their limit for $q_{1.4}$ we find that only 2 of the COLA galaxies, IRAS 14566-1629 and IRAS 13197-1627, qualify as ``radio AGN'' (1.9\% of our sample, in good agreement with Yun et al.), both of which have compact radio cores and are optically classified as AGN. However, our study indicates that this selection criterion under-estimates the number of AGN with detectable compact cores. If we consider only those galaxies in the COLA sample which have compact cores and are classified optically as Seyferts, their highest value of $q_{1.4} = 2.16$. A total of 12 galaxies in the COLA sample have $q_{1.4} \le 2.16$ of which half have compact radio cores and are optically classified as Seyferts. Of the remaining 6 galaxies, one is a Seyfert galaxy (Kewley et al. 2001) without a compact radio core ($<1.5mJy$), two are HII galaxies (Corbett et al, in preparation) and three galaxies have yet to be classified.  Our results therefore suggest that defining ``radio AGN'' as galaxies whose radio luminosity is 2.5 times that predicted from the FIR emission would still yield a high proportion of AGN; 5/7 galaxies (71\%) of the COLA galaxies with radio fluxes more than 2.5 times that predicted have compact radio cores and are Seyferts. Compact cores were not detected in the remaining two galaxies but since they have yet to be classified optically, we cannot rule out the possibility that these too are AGN.    
  
As well as the radio excess exhibited by the galaxies with compact cores it is 
interesting to look for trends within the data on compact cores (although with 
only 9 objects we must be wary of small number statistics). We find that the luminosity of the compact cores is correlated with the total radio 
luminosity (Fig. 6a). This result is significant at $>$99.5\% level (Kendall $\tau$ and Spearman $\rho$ tests) for the 4.8 GHz and 1.38 GHz data and $>95.0\%$ for the 2.5 GHz data. The correlation between the core luminosity and the total radio luminosity is not surprising given the radio excess exhibited by all but one of the galaxies with compact cores. If the radio excess is due to structures associated with the compact core, then one might expect that the radio emission of these structures would be related to the radio luminosity of the compact core and since they contribute a large proportion of the total radio emission of the galaxies, the total radio luminosity of the galaxies will be dependent on the luminosity of the compact core. 
    
The luminosity of the compact cores is independent of the FIR luminosity (Fig. 6b) indicating that the compact core (and any structures associated with it) contributes little to the overall FIR emission of the galaxy. 

Finally, the luminosity of the compact cores is inversely proportional to $q$ (or proportional to the radio excess) with a formal significance of $>95\%$ measured using Kendall's $\tau$ test and Rank Spearman's test for the 4.8 GHz data. This result is not surprising given the correlations described in the previous paragraphs. If the core luminosity is proportional to the total 
radio luminosity and independent of the FIR luminosity, it will be inversely 
proportional to the ratio of the FIR and radio luminosities.    
  
Detailed notes on the objects with compact cores are given in Appendix A.

\subsection{Compact cores - radio supernovae or AGN?} 

An 18cm VLBI survey of 31 luminous ($L_{FIR} > 10^{11.25}L_{\sun}$) galaxies 
conducted by Smith, Lonsdale \& Lonsdale (1998) detected compact (5-150 milli-
arcsec) structures in 21 objects. Of the objects with compact cores, optical 
spectra obtained for 19 of the galaxies indicated that roughly half were 
optically classified as AGN (the majority LINERS) and the rest HII/starburst 
galaxies. However we note that 8 of the galaxies with compact cores lie very 
close (within $\sim1$dex) to the partition between AGN and HII regions (3 on the AGN side, 4 on the HII region side and 1 on the partition). The optical 
classification of these objects is therefore by no means secure. Although Smith, Lonsdale \& Lonsdale concluded that all the compact cores could be AGN, they 
also found that many (but not all) of the compact cores could also be attributed to clumps (2-16pc in size) of starburst-generated radio supernovae (RSN).

Similar results were obtained by Kewley et al. (2000a) from their survey of 48 
galaxies with the PTI and they were unable to rule out starburst-generated RSN 
complexes in all but 5 of the 22 galaxies in which they detected compact cores. 
However, unlike Smith, Lonsdale \& Lonsdale (1998), they find a relationship 
between the optical classification and the luminosity of the compact core, with 
the galaxies classified as HII or starburst galaxies (11 from 22) having lower 
luminosities than the galaxies classified as AGN (9 from 22).  

Several models exist for predicting the star formation rate (SFR) and consequent supernovae rate and radio emission from the FIR luminosity of the sources (see for example Smith, Lonsdale \& Lonsdale 1998; Kewley et al. 2000a). However it is not clear that these models adequately describe the extreme conditions that must exist in these supernovae complexes. Additionally, since these models use the FIR luminosity to predict the star formation rate, we would expect the luminosity of the radio cores to scale with the FIR luminosity of the host galaxies, which, as discussed in the previous section, is not the case. 
    
Clearly the compact core we detect in the 5 galaxies classified as Seyferts is  most likely to be the AGN. The situation is, however, not so clear for the other 3 objects. If we consider the luminosity of the compact cores in the borderline objects IRAS 09375-6951 and IRAS 13097-1531 ($10^{4.2} L_{\sun}$ and $10^{4.1} L_{\sun}$) we conclude that these objects probably contain AGN, but the optical spectrum is dominated by star formation activity in the host galaxy. Finally, the starburst object, IRAS 13135-3801, exhibits a compact core of $10^{3.7}L_{\sun}$, well below the $10^{4} L_{\sun}$ upper limit found by Kewley et al. 2000a) for their SNR candidates, and no radio excess. We therefore conclude that the compact core detected in this galaxy is more likely to be a complex of supernovae remnants than a low luminosity AGN which is dominated at optical to radio wavelengths by the starburst activity in the host galaxy.

\subsection{Non-detections of galaxies with previously detected radio cores}

It is also interesting to look at the objects in which we {\it failed} to detect compact radio cores. We failed to detect compact cores with the ATCA-Tidbinbilla baseline in three galaxies in which cores were detected by Kewley et al. (2000) 
using the PTI. The three galaxies are IRAS 04118-3207 ($<$1.4mJy), IRAS 10221-
2317 ($<$2.5mJy) and IRAS 12596-1529 ($<$2.1mJy). Kewley et al. (2000a) found these galaxies to have cores of 1mJy, 2mJy and 2mJy respectively. None of these 
galaxies exhibit a significant radio excess, having values of $q_{4.8} =$ 2.87, 2.80 and 2.89 respectively. Both IRAS 04118-3207 and IRAS 10221-2317 are 
optically classified as borderline or transition objects, (Kewley et al. 2000a; 
our optical spectra) while IRAS 12596-1529 is classified as a starburst or HII-
like region (Kewley et al. 2000a). 

Given that our upper detection limits for these objects are slightly higher than the flux measured by Kewley et al. (2000a), it is not surprising that we do not 
detect them. All three galaxies have compact cores $<10^{3.7} L_{\sun}$ and were considered candidates for complexes of SNR rather than AGN. 

Compact radio cores are also not detected in several galaxies previously 
classified in the literature as AGN. It is difficult to compare literature 
classifications directly as the slit width, extraction window and slit 
position often vary from observer to observer. Additionally, objects may be classified using only the $[OIII]/H\beta$ and $[NII]/H\alpha$ line ratios. A galaxy may exhibit AGN-like emission line ratios when these lines are compared but starburst-like emission line ratios when the $[SII]/H\alpha$ and $[0I]/H\alpha$ ratios are examined. These slight differences, while not altering the classifications of the galaxies which exhibit strong AGN or starburst emission line ratios, do result in border -line objects being classified differently by different authors.  For example, IRAS 04118-3207 has been classified as a Seyfert 2 (Maia et al. 1996) on the basis of its $[OIII]/H\beta$ and $[NII]/H\alpha$ line ratios but is classified as a 
starburst galaxy by Kewley et al. (2000a) using the [OI] and [SII] emission line ratios. Kewley et al.'s final classification for IRAS 04118-3207 is a borderline object. We have therefore decided not to use the classifications of the galaxies in our sample in the literature and obtained optical spectra for all the galaxies in this paper. The optical classification of the galaxies and its relationship to the radio luminosity and FIR colors of the galaxies will be discussed in Corbett et al., in preparation.    
 
\section{Conclusions}

This paper represents the first in a series of papers which discuss the multi-
wavelength properties of a large sample of galaxies. When we have completed our 
survey we will select three sub-samples of galaxies, AGN with compact cores, AGN without compact cores and galaxies without AGN, for comparison of their 
circumnuclear regions. The subsamples will be matched in such orientation - 
independent properties as the total radio luminosity and the CO mass as 
determined from the initial survey.

Here we present measurements of the total radio luminosity of the Southern COLA 
sample at 4.8, 2.5 and 1.4  GHz. Of the 107 galaxies observed, fluxes were 
measured for 104 galaxies at 4.8 GHz, 94 at 2.5 GHz and 84 at 1.4 GHz. 
Australian LBA snapshot observations were obtained for 105 of the galaxies at 
2.38 GHz and 0.05-0.08 arcsec resolution. Compact cores ($<$0.05 arcsec) were 
detected in 9 galaxies with fluxes $>1.6 $mJy and brightness temperatures 
$T_{B}>10^{5.3}K$. The luminosity of the compact cores varied between 
$10^{3.4}L_{\sun}$ - $10^{5.6}L_{\sun}$. 

We find the following:
\begin{itemize} 
\item The majority of COLA galaxies follow well known radio-FIR correlation at 
all three radio frequencies with the relationship scaling as a function of 
wavelength. They exhibit a slight FIR excess compared to observations of spirals or normal galaxies, possibly due to an excess of hot dust or gas or to selection effects. There is no evidence that the FIR excess increases with the FIR luminosity of the galaxies.

\item As a whole, the nine galaxies with compact cores exhibit a radio excess 
relative to the other galaxies in the sample. Six of the galaxies with compact cores exhibit a radio excess of more than $1\sigma$ at all radio frequencies and only one galaxy with a compact core exhibits a radio deficit. This excess is statistically significant (at the 99.9\% level) and remains after the emission from the compact core is subtracted,  implying that there are radio structures associated with the cores extending further out than 0.05 arcsec, 15-25 pc at the COLA galaxy redshifts.

\item There is some evidence that the luminosity of the compact cores is 
dependent on the total radio luminosity of galaxies and the radio excess 
exhibited by galaxies, i.e. galaxies with a large radio excess tend to have 
higher luminosity compact cores. The luminosity of the compact cores is 
independent of the FIR luminosity, indicating that the activity associated with 
the compact cores does not contribute significantly to the overall FIR luminosity of the galaxy.

\item From the optical emission line ratios, six of the galaxies with compact 
cores are classified as Seyferts, two are borderline AGN/HII (or transition) 
galaxies and one galaxy is a starburst. We find evidence that the optical classification depends on the radio excess exhibited by the galaxies with the six galaxies classified as Seyferts exhibiting the largest radio excesses, the two borderline objects exhibiting smaller radio excesses and the starburst galaxy exhibiting a radio deficit.  This is presumably due to the fact that the radio excess may in fact measure the relative contributions of star-formation and AGN emission to the observed luminosity.  

\item From consideration of the radio properties and optical emission-line ratios we believe that the source of the compact radio emission in 8/9 objects is an AGN  whereas it is plausible that a complex of compact 
supernovae remnants as seen in, e.g. Arp 220 (Smith, Lonsdale \& Lonsdale 1998) 
is responsible for the compact core in one galaxy, IRAS 13135-2801. This conclusion can be tested by obtaining high resolution (pc-scale) radio images of the galaxies, since any highly collimated structures, such as the jet observed in one of the Seyfert galaxies, IRAS 13197-1531, are indicative of AGN rather than starburst activity. Additional evidence could come from X-ray observations since AGN exhibit a hard X-ray spectra whereas the X-ray spectra due to starburst activity is soft.

\end{itemize}
   
Our results indicate that a large $>30\%$ radio excess relative to the radio-FIR correlation is strongly suggestive of the presence of a compact core in a galaxy and this property could be used in future surveys to select candidate galaxies 
for compact cores. Of the 14 galaxies with $q_{4.8}$ more than $1\sigma$ lower 
than $\langle q_{4.8} \rangle$, half have compact radio cores. The proportion of AGN increases with the radio excess and of those galaxies whose radio luminosity is 2.5 times that predicted from the FIR emission 71\% (5/7) have compact cores and are optically classified as AGN. The remaining two galaxies have yet to be classified optically but do not possess compact cores. Surveys which rely upon spectral indices alone for selecting radio AGNs may significantly underestimate the number of galaxies containing AGN cores because of possible contamination of the overall spectral index by a starburst component. 

The compact cores themselves are too weak to account for the observed radio excess and this in turn implies that the galaxies optically classified as AGN have associated radio structures on scales $>15$pc which cause them to depart from the FIR-radio correlation. Studies by Baum et al. (1993) and Roy et al. (1998) imply that these structures could persist out to $\sim$1kpc and at least one of the objects with a compact 
core, IRAS 13197-1627, is known to possess a $>$ 250pc scale radio jet (Kinney 
et al. 2000). It is therefore highly likely that structures analogous to the 
jets and radio lobes seen in radio loud quasars are responsible for this radio 
excess. It is interesting to note, however, that the compact cores contribute 
little to the FIR luminosity of the galaxies.    

For the COLA project these results are very encouraging. There is strong 
evidence that that 8/9 of the compact cores we detect are AGN, there is no 
evidence of any orientation biases in our detection of compact cores and the AGN dominates the starburst emission at both radio and optical wavelengths in 6/9 of the galaxies.  These 8 AGN can therefore form the Southern sample of 
AGN galaxies with compact cores in our study of the circumnuclear regions of 
host galaxies.

When we have completed the optical classification of the southern COLA galaxies, we will be in a position to select two further samples of galaxies - AGN without compact cores and galaxies without AGN. These galaxies will be matched to the 
AGN with compact cores using properties of the host galaxies such as the CO 
mass, the I-Band luminosity and the FIR luminosity. High resolution ATCA, LBA 
and NIR observations will be combined with optical spectroscopic and emission 
line imaging. These observations will enable us to investigate the internal 
kinematic properties of their gas and the spatial distribution of emission 
mechanisms such as photoionisation and shock-excitation, e.g. ionisation cones. 
We also aim to establish whether their radio structure at pc-and kpc- scales is 
aligned or associated with any optical structures.

From studies such as Roy et al. (1998) we expect that the optical classification of the southern COLA galaxies will reveal a population of objects with optical 
AGN classifications which do not exhibit a significant radio excess and have no 
detected compact radio core. While it is reasonable that many of these objects 
do possess low luminosity compact cores below our detection limit they also seem to be missing the associated larger scale radio structures seen in many of the 
objects with compact cores.  It is unlikely that the absence of a radio excess 
in these galaxies is simply due to a larger contribution to the overall radio 
emission from star-formation, since this would also reduce the strengths of the 
AGN emission-line signature in the optical spectrum, moving the objects towards 
the borderline regions on the diagnostic diagrams. It may be that the galaxies 
with compact cores are simply ``more powerful" AGN, perhaps possessing larger 
black holes or a different nuclear environment and/or accretion rates relative 
to the AGN without compact cores. Alternatively it could be that the Seyfert 
galaxies with compact cores represent low luminosity, non-relativistic, small 
scale versions of the much larger and more luminous quasars, blazars and radio 
galaxies whereas the AGN without compact cores simply do not possess the same 
collimated jets. In addition to providing important information on the 
differences (if any) between the circumnuclear regions of AGN and `non-AGN' 
galaxies, the COLA project may also shed light on radio-loud/radio-quiet 
dichotomy.          

\section*{Acknowledgements}
This project is dedicated to our dear friend and colleague, Charlene Heisler, who sadly passed away in October 1999. She was one of the founders of the COLA project and her enthusiasm and scientific insight will be sorely missed.

The authors would like to thank all those who have helped with this project: Lisa Kewley who provided her spectroscopic data reduction script and allowed us to access to her results before they were published; I. Wormleaton for her help in reducing the optical spectra and the ATNF LBA team who provided a great deal of help at every stage of our LBA observations. We also thank Geraint Lewis for his careful reading of the final manuscript. The observations presented in this paper were obtained on the Australia Telescope Compact Array, Parkes 64m Telescope and Tidbinbilla DSN telescope. The Australia Telescope is funded by the Commonwealth of Australia and operated as a National facility, managed by CSIRO. Observations from the Australian National University 2.3m Telescope at Siding Springs Observatory, NSW were also discussed. This research made use of the NASA/IPAC Extragalactic Database (NED) which is operated by the Jet Propulsion Laboratory, California Institute of Technology under contract with NASA. Part of the research was conducted while EAC and MAD were the recipients of an International Research Exchange Fellowship, funded by DEETYA. EAC also acknowledges an ATNF Visting Fellowship. We acknowledge the support of NSF grant INT 9418142 for travel funds
for the US collaborators in this project.

\appendix
\section{Notes on individual galaxies}
  
The nine galaxies in which we detected compact radio cores are discussed in more 
detail below.
\begin{itemize} 
\item {\bf IRAS 09375-6951} is classified in the NASA Extragalactic Database 
(NED) as an irregular spiral galaxy. It was observed twice with the LBA, once 
with the ATCA-Tidbinbilla LBA (1998 September) and once with the ATCA-Parkes LBA (2000 July). A compact core was detected with a flux of 3.6mJy ($10^{4.25}L_{\sun}$) on the ATCA-Tidbinbilla baseline and a flux of 4.4mJy on the Parkes-ATCA baseline. It is possible that this difference in flux reflects intrinsic variability of $\sim$ 20\% in the core emission since the observations were separated by 22 months but it is more likely that it is due to the different lengths of the two baselines. The resolution of the Parkes-ATCA LBA (0.08 arcsec) is lower than that of the ATCA-Tidbinbilla (0.05 arcsec) and thus will contain radio emission from a larger area of the source. The 0.05 arcsec compact radio core contributes 13\% of the total radio emission and the radio excess at 4.8 GHz is 0.31. No published optical spectra are available but optical spectra we have obtained show it to be a borderline or transition object, intermediate between AGN and starburst (e.g. Hill et al. 1999, Barth \& Shields 2000). 

IRAS 09375-6951 could either contain an AGN which is obscured at optical 
wavelengths or a complex of supernovae remnants (e.g. M82; Unger et al. 1984). 
As discussed in Section 2.2.2,  Kewley et al. (2000a) found that the sources they believed to be complexes of SNR tended to have compact cores with luminosities of $<10^{4}L_{\sun}$. The 0.05 arcsec compact core in IRAS 09375- 6951 has a luminosity of $\sim 10^{4.3}L_{\sun}$ which would place it among the AGN rather than the starburst galaxies.  

\item {\bf 12329-3938} (NGC 4507) is a barred spiral galaxy (SAB(s)ab; Sandage \& Brucato 
1979) and has been classified as either a Sy 2 (Durret \& Bregeron 1986) or a Sy 1.9 (Veron-Cetty \& Veron 1998) indicating that broad H$\alpha$ but not broad 
H$\beta$ is sometimes visible in the spectrum. Our optical spectra also show it to be a Sy2 with similar line ratios to the other AGN in 
our sample. This source was observed with the PTI by Roy et al. (1998) but no 
core was detected at the level of 5mJy. We make a tentative detection of a 
$<0.05$ arcsec core at 1.6mJy ($10^{3.4}$ L$\sun$), making this 
source the lowest luminosity core detection in our sample. The detection is 
tentative as the flux in the fringe-rate feature is at a similar level to the 
noise. However, both the right and left polarisation data show exactly the same 
feature at the same fringe rate, which is unlikely to be a coincidence.  

IRAS 12329-3938 exhibits a large radio excess (0.85) which remains after 
subtraction of the compact core and this, combined with the optical AGN 
classification, is strong evidence that the compact core we detect is a weak AGN rather than a supernovae complex.   

\item {\bf IRAS 13097-1531} (NGC 1510) This object is a S0 (or peculiar spiral) and is 
known to contain an OH megamaser (Martin et al. 1989). From optical 
spectroscopic observations it has been classified as a ``composite object'' by 
Baan, Saltzer \& Le Winter (1998) and our own observations show that it lies close to the AGN/HII region partition. We detect a compact core at the level of 2.5mJy. This object exhibits a small radio excess,  0.27, at 4.8GHz and one of the faintest compact cores. IRAS 13097-1531 is similar to IRAS 09375-6951 in that the optical classification is ambiguous but the radio luminosity of the core is $10^{4.14}$ L$\sun$, brighter than would be expected for the supernova complexes.

\item {\bf IRAS 13135-2801} (NGC 5051) is an SA(rs)b galaxy with a nuclear ring (Buta 
1995). We detect a compact core in both polarisations at the level of 2.2mJy 
($10^{3.75}$ L$\sun$). This is the only object with a compact radio core which 
exhibits a {\it deficit} in radio flux, emitting 17\% {\it less} radio flux than would be predicted from the FIR luminosity. Optical spectra we have obtained show it to be a starburst or HII galaxy. The low luminosity of the 
radio core combined with the optical spectroscopic classification means that it 
is highly possible that we have detected a complex of radio supernovae in this 
object rather than an AGN (see Section 4.6 for futher discussion). 

\item {\bf IRAS 13197-1627} is a known Seyfert galaxy (de Grijp et al. 1985) and is classified as a Seyfert 1.8 by Aguerro et al. (1994). It has a tentative 
classification in NED as an SB galaxy morphology. A compact core of 26mJy has 
been detected previously using the PTI (resolution 0.1 arcsec; Heisler, Lumsden 
\& Bailey 1997) and VLA observations at 3.6cm by Kinney et al. (2000) have 
revealed a linear structure extending about 278pc from the core, believed to be a synchrotron jet. We detect a compact core of 8.9 mJy at 0.05 arcsec 
resolution. IRAS 13197-1627 exhibits one of the largest radio excesses of our 
sample with the 4.8 GHz flux almost an order of magnitude larger than the 
predicted value. This excess remains after subtraction of the compact core 
emission and is probably due, at least in part, to emission from the $\sim$ 300pc-scale radio jet. 

\item {\bf IRAS 14544-4255} (IC 4518A) is one of a pair of strongly interacting 
galaxies (Condon et al. 1996) but is not as yet classified as an AGN. The total 
radio luminosity at 4.8 GHz is 2.5 times that predicted from the FIR luminosity. 
The compact radio core (3.77 mJy) contributes only 3.5\% of the total radio 
emission but even after subtraction of the core emission it still exhibits twice as much radio emission as would be expected from the FIR emission. No published 
optical spectra are available but optical spectra we have obtained (Corbett et 
al. 2000) show it to be a Sy 2. 

\item{\bf IRAS 14566-1629} (NGC 5793) has an Sb classification (NED) and is one of a pair 
of galaxies. VLBI observations of this source have detected a compact nucleus 
(0.016 arcsec) which is extended parallel to the minor axis of the galaxy 
(Gardner et al. 1992). There is also some evidence for a water maser with a 
diameter larger than 20pc (Hagiwara et al. 1997) and it was recently classified 
as a Seyfert 2 galaxy both by us and Baan, Saltzer \& Le 
Winter (1998). IRAS 14566-1629 exhibits the largest radio excess of the galaxies in our sample with a radio luminosity at 4.8 GHz more than 30 times that 
expected from the FIR luminosity and the brightest compact core (266mJy).  This 
core accounts for 30\% of the observed radio flux.
  
\item {\bf IRAS 19543-3804} is classified as a spiral galaxy (Sb or IrS) in a 
cluster. No published optical spectra is available but spectra we 
have obtained show it to be a Seyfert 2.  It was observed 
twice with the LBA, once with the ATCA-Tidbinbilla LBA (September 1998) and once with the ATCA-Parkes LBA (July 2000). A compact core was detected with a flux of 23.4mJy ($10^{5.0}L_{\sun}$) on the ATCA-Tidbinbilla baseline and a flux of 24.7mJy on the Parkes-ATCA baseline. Although we cannot rule out the possibility that the difference between the fluxes is due to $\sim$5\% variability of the source, it is more likely that the a small percentage of the core flux measured on the ATCA-Parkes baseline is resolved on the Tidbinbilla-ATCA baseline.    

The 0.05 arcsec compact radio core at $10^{4.9}$ L$\sun$ is one of the brightest we detect and accounts for 32\% of the total radio luminosity at 2.3 GHz. This 
source exhibits a large radio excess, emitting more than 5 times the radio 
emission at 4.8 GHz than would be expected from the FIR luminosity. Subtraction 
of the compact core reduces the radio excess to $\sim$ 40\%.  

\item {\bf IRAS 21453-3511} (NCG7130, IC 5135)) is a known AGN, identified by Philips et al. (1983) 
as a Seyfert 2 active nucleus surrounded by a starburst ring. Observations by 
Thuan (1984) further showed that the starburst dominates the galaxy emission 
contributing 75\% of the emission in the UV. Its optical emission line ratios 
have confirmed its AGN classification but it falls on the borderline between the Sy2 and LINER classes with Veilleux et al. (1995) classifying it as a LINER while 
Heisler, Lumsden \& Bailey  (1997) and Vaceli et al. (1997) classify it as a 
Seyfert 2. We also classify it as a Seyfert 2 and it is possible that the differences between the classifications reflect differing 
contributions from the starburst emission in the slit. 

We detect a 4.7mJy ($10^{4.1}$ $L_{\sun}$) core in IRAS 21453-3511 and a radio 
excess at 4.8 GHz of 92\% (i.e. the 4.8 GHz luminosity is nearly twice that 
predicted from the FIR luminosity). The compact radio core contributes 3.8\% of 
the total radio luminosity at 2.3 GHz but after subtraction of the core the 
source still exhibits an excess radio flux above the radio-FIR correlation of 
$>$80\% of the total radio luminosity.  Heisler et al. (1997) measured a compact core flux at 2.3 GHz of 14mJy using the PTI baseline with a resolution of 0.1 
arcsec, twice that of ours, which implies that the source has compact radio 
structures between 25-50pc in size which we have resolved out. These structures 
could be associated with a radio jet. Alternatively, the difference between the PTI flux and our measurements could be due to intrinsic variability of the compact core. 

\end{itemize}

\begin{deluxetable}{llcccccccc}
\tabletypesize{\scriptsize}
\tablewidth{0pc}
\tablecolumns{14}
\tablecaption{Sources in the Southern COLA sample and their ATCA and LBA flux measurements.}
\tablecomments{The sources in the COLA sample are indentified by their IRAS name, with their more common name given in the next column. The heliocentric velocities are quoted are from Strauss et al. (1992) and have typical errors of 1\% or less. The RA and Dec given is the centroid of the fit to the 4.8GHz data, with the exception of the three sources not detected at 4.8GHz for which we quote the RA and Dec from the IRAS point Source Catalogue. The errors in the centroid position are typically $<$1arcsec.  All fluxes are quoted in mJy and the ATCA flux measurements have typical errors of 10\% at 4.8 and 2.5GHz and 15\% at 1.4. ND=Source not detected with flux greater that 5$\times$ the rms of the map, C= confusion i.e. the beam overlapped two objects, only one of which was the target source. All LBA fluxes are given in mJy and were measured with the ATCA and Tidbinbilla baseline except those marked with an asterisk (*) which were observed with the ATCA and Parkes telecsope only. Those marked A were observed in 1998 September, while those obtained in 2000 July are marked B. For those sources detected with LBA the rms variation in the visibility is given in Table 4 and we estimate that there may be a systematic error in the flux of 1-2\% from the method used to detrmine the antennae gains. In the final column we indicate whether the source was extended (E) or point-like (P) in the 4.8GHz maps. }
\tablehead{
\colhead{IRAS} &  \colhead{Alternative}     & \colhead{Helio Vel}& \colhead{RA(2000)}      &  \colhead{Dec(2000)}   & \colhead{4.8GHz}   &  \colhead{2.5GHz} &  \colhead{1.38GHz} & \colhead{LBA} & \colhead{4.8GHz}\\
\colhead{ Name}             & \colhead{Name}    &   \colhead{(km s$^{-1}$)}  & \colhead{(hh mm ss) } & \colhead{ (\degr  '  ") } & \colhead{Flux}& \colhead{Flux}  &  \colhead{Flux}& \colhead{Flux} & \colhead{Morph.}\\}
\startdata
00085-1223& N 34        & 5821& 00 11  6.53& -12 06 26.6& 29.6& 44.6& 58& $<$ 2.3$^{B}$& P\\
00344-3349& 0034-3349   & 6156& 00 36 52.46& -33 33 16.9& 15.3& 15.5& ND   & $<$ 1.5$^{A}$& P\\
00402-2350& N 232 M     & 6647& 00 42 45.83& -23 33 40.7& 22.8& 35.9& 63& $<$ 2.6$^{B*}$& P\\
01053-1746& I1623       & 6016& 01 07 47.21& -17 30 23.9& 74.6& 146.0& 211& $<$ 1.7$^{A}$& E\\
01159-4443& 0115-4443BM & 6701& 01 18  8.35& -44 27 43.9& 21.5& 25.8& 41& $<$ 1.4$^{A}$& P\\
01165-1719& I 93        & 5977& 01 19  2.21& -17 03 36.3& 11.3& ND   & 29& $<$ 1.7$^{B}$& P\\
01326-3623& 0132-3623   & 4827& 01 34 51.26& -36 08 14.2& 30.7& 52.4& 56& $<$ 1.9$^{A}$& P\\
01341-3734& N 633       & 5180& 01 36 23.51& -37 19 22.0& 10.4& 25.2& ND   & $<$ 1.8$^{A}$& E\\
01384-7515& N 643B      & 3966& 01 39 13.33& -75 00 39.1& 15.6& 25.1& 38& $<$ 1.1$^{A}$& P\\
02015-2333& 0201-2333   & 4934& 02 03 56.68& -23 18 46.6& 9.8 & 23.8& 22& $<$ 2.7$^{B*}$& E\\
02069-1022& N 835 M     & 4152& 02 09 24.53& -10 08 11.0& 12.5& ND   & ND   & $<$ 1.9$^{B}$& P\\
02072-1025& N 839       & 3847& 02 09 42.77& -10 11  4.5& 19.5& ND   & ND   & $<$ 1.6$^{B}$& P\\
02140-1134& N 873       & 4009& 02 16 32.34& -11 20 49.6& 15.9& 36.2& 53& $<$ 2.2$^{B}$& E\\
02281-0309& N 958       & 5738& 02 30 42.83& -02 56 24.8& 20.3& 20.2& C    & $<$ 1.9$^{B}$& P\\
02433-1534& N1083       & 5507& 02 45 40.56& -15 21 32.3& 20.2& 35.9& 50& $<$ 1.8$^{B}$& P\\
02436-5556& 0243-5557   & 4101& 02 45  8.82& -55 44 25.5& 11.1& 16.9& ND   & $<$ 1.8$^{A}$& P\\
02476-3858& 0247-3858   & 5008& 02 49 34.23& -38 46 13.0& 6.3 & ND   & ND   & $<$ 1.8$^{A}$& P\\
03022-1232& N1204       & 4282& 03 04 39.90& -12 20 28.6& 14.0& 16.4& 12& $<$ 1.6$^{B}$& P\\
04118-3207& 0411-3207   & 3570& 04 13 49.66& -32 00 25.4& 26.0& 40.7& 61& $<$ 2.2$^{A,B*}$& P\\
04210-4042& N1572       & 6012& 04 22 42.82& -40 36  3.1& 18.3& 34.1& 44& $<$ 1.9$^{A}$& P\\
04315-0840& N1614       & 4744& 04 34  0.04& -08 34 47.1& 63.4& 86.2& 124& $<$ 2.8$^{B*}$& P\\
04335-2514& 0433-2514   & 4843& 04 35 39.28& -25 07 59.2& 15.9& 22.7& 12& $<$ 2.5$^{B*}$& P\\
04370-2416& 0437-2416   & 4422& 04 39  6.33& -24 11  2.5& 15.7& 23.0& 52& $<$ 1.5$^{A}$& P\\
04461-0624& N1667       & 4578& 04 48 37.09& -06 19 15.2& 23.7& 56.6& 55& $<$ 1.8$^{B}$& E\\
04501-3304& 0450-3304   & 5622& 04 52  4.79& -32 59 26.1& 9.1 & 10.1& 11& $<$ 1.6$^{A}$& P\\
04558-0751& 0455-0751   & 3773& 04 58 12.53& -07 46 52.2& 13.4& 50.2& 82& $<$ 1.4$^{B}$& P\\
04569-0756& N1720       & 4242& 04 59 20.61& -07 51 30.7& 14.3& 17.8& 30& $<$ 2.7$^{B*}$& P\\
04591-0419& N1741C      & 4068& 05 01 37.83& -04 15 16.4& 9.7 & C    & C    & $<$ 1.5$^{B}$& E\\
04595-1813& N1738 M     & 3978& 05 01 46.80& -18 09 18.7& 11.7& 55.0& 100& $<$ 2.3$^{B*}$& E\\
05041-4938& N1803       & 4145& 05 05 27.07& -49 34 12.3& 2.8 & 4.0& 24& $<$ 1.6$^{A}$& P\\
05053-0805& N1797       & 4478& 05 07 44.91& -08 01 7.0& 10.7& 15.0& 32&  Nobs&  P\\
05140-6213& 0514-6213   & 4966& 05 14 29.66& -62 10 16.0& 3.8 & 11.3& 11& $<$ 1.7$^{A}$& P\\
05449-0651& A0544-0651  & 6467& 05 47 21.1  & -06 50 17   & ND   & ND   & ND   & $<$ 1.7$^{B}$& \\
05562-6933& N2150       & 4440& 05 55 46.26& -69 33 38.8& 10.7& 20.1& 26& $<$ 1.7$^{A}$& P\\
06295-1735& 0629-1735   & 6339& 06 31 47.21& -17 37 16.2& 9.0 & 18.2& 35& $<$ 1.5$^{B}$& P\\
06592-6313& 0659-6313   & 6882& 06 59 40.37& -63 17 53.3& 8.6 & 12.0& 9& $<$ 2.6$^{B*}$& P\\
08175-1433& 0817-1433   & 5732& 08 19 49.58& -14 42 58.3& 16.0& 32.4& 54& $<$ 1.3$^{B}$& P\\
08225-6936& 0822-6936   & 3924& 08 22 40.29& -69 46 20.4& 8.9 & 11.1& 14& $<$ 1.7$^{A}$& P\\
08364-1430& 0836-1430   & 4184& 08 38 46.03& -14 40 52.2& 12.4& 19.5& 28& $<$ 1.3$^{B}$& E\\
08438-1510& 0843-1510   & 5423& 08 46  9.30& -15 21 27.4& 9.0 & 15.2& 19& $<$ 2.1$^{B}$& P\\
09006-6404& 0900-6404   & 6636& 09 01 37.15& -64 16 25.6& 11.5& 16.4& 25& $<$ 1.6$^{A}$& P\\
09248-1918& 0924-1918   & 4888& 09 27 13.43& -19 31 53.2& 9.9 & 15.5& 14& $<$ 2.5$^{B*}$& P\\
09375-6951& 0937-6951   & 6066& 09 38 19.31& -70 05 28.0& 17.4& 26.6& 41&   3.6$^{A,B*}$& P\\
10015-0614& N3110       & 5034& 10 04 02.08& -06 28 36.6& 38.0& 76.3& 96& $<$ 1.6$^{B}$& E\\
10221-2317& 1022-2317   & 3662& 10 24 31.48& -23 33 11.1& 22.7& 38.4& 39& $<$ 2.7$^{B*}$& P\\
10409-4557& 1040-4556   & 7000& 10 43  8.00& -46 12 47.2& 13.4& C    & C    & $<$ 1.7$^{A}$& E\\
10484-0153& I 651       & 4464& 10 50 58.3  & -02 09 00   & ND   & 24.3& 30& $<$ 1.3$^{B}$& \\
10567-4310& 1056-4310   & 5156& 10 59  1.75& -43 26 25.1& 11.3& 19.6& 32& $<$ 1.6$^{A}$& P\\
11005-1601& N3508       & 3877& 11 02 59.75& -16 17 20.9& 17.2& 45.2& 59& $<$ 2.6$^{B*}$& E\\
11254-4120& 1125-4120   & 4902& 11 27 54.11& -41 36 51.7& 10.3& 14.4& 21& $<$ 1.3$^{A}$& P\\
11328-4844& 1132-4844   & 5624& 11 35 16.24& -49 00 39.6& 13.2& 21.4& 24& $<$ 1.6$^{A}$& P\\
11409-1631& 1140-1631   & 3660& 11 43 29.69& -16 47 42.0& 8.2 & 17.8& 17& $<$ 1.6$^{B}$& P\\
12042-3140& 1204-3140 M & 6818& 12 06 51.88& -31 56 59.2& 19.1& 30.1& 43& $<$ 1.7$^{A}$& P\\
12112-4659& 1211-4659   & 5493& 12 13 51.90& -47 16 23.1& 7.1 &  9.2& 20& $<$ 1.5$^{A}$& P\\
12115-4657& 1211-4657   & 5543& 12 14 12.81& -47 13 43.9& 18.1& 12.1& 38& $<$ 1.7$^{A}$& P\\
12120-1118& 1212-1118   & 5406& 12 14 41.99& -11 34 57.6& 11.7& 23.1& 30& $<$ 1.6$^{B}$& E\\
12171-1156& 1217-1156   & 4234& 12 19 42.26& -12 13 31.1& 5.1 & 12.8& 15& $<$ 1.9$^{B}$& P\\
12286-2600& 1228-2600   & 5970& 12 31 16.97& -26 17 14.7& 6.6 & ND   & ND   & $<$ 1.4$^{B}$& P\\
12329-3938& N4507       & 3523& 12 35 36.52& -39 54 32.7& 17.7& 38.89& 57&   1.6$^{A,B}$& E\\
12351-4015& N4575       & 5250& 12 37 51.40& -40 32 15.1& 16.0& 29.6& 49& $<$ 2.0$^{A}$& E\\
12504-2711& 1250-2711   & 3694& 12 53 11.07& -27 27 52.0& 11.1& ND   & 44& $<$ 1.6$^{B}$& E\\
12596-1529& 1259-1529   & 4773& 13 02 19.92& -15 46  2.9& 12.4& 24.1& 29& $<$ 1.6$^{B}$& P\\
13001-2339& 1300-2339   & 6446& 13 02 52.36& -23 55 16.5& 36.8& 50.0& 85& $<$ 1.4$^{B}$& P\\
13035-4008& 1303-4008   & 4475& 13 06 25.94& -40 24 52.0& 12.6& 19.2& 24& $<$ 1.3$^{A}$& P\\
13097-1531& N5010       & 6400& 13 12 26.35& -15 47 51.6& 29.3& 51.6& 88&   2.5$^{B}$& E\\
13135-2801& N5051       & 4426& 13 16 20.17& -28 17  9.8& 8.8 & 16.7& 21&   2.2$^{B}$& P\\
13193-5208& 1319-5208   & 5090& 13 22 22.39& -52 23 45.3& 5.1 &  7.7& ND   & $<$ 2.0$^{A}$& P\\
13197-1627& 1319-1628   & 5152& 13 22 24.47& -16 43 42.0& 97.9& 181.3 & 259 &   8.9$^{B}$& E\\
13229-2934& N5135       & 4112& 13 25 44.00.& -29 50  1.2& 69.5& 119.6& 197& $<$ 1.5$^{B}$& P\\
13301-2357& I4280       & 4889& 13 32 53.27& -24 12 25.1& 13.3& 31.9& 61& $<$ 1.8$^{B}$& P\\
13322-3500& 1332-3500   & 3952& 13 35  4.74& -35 16  8.8& 15.4& 28.5& 35& $<$ 1.9$^{A}$& P\\
14132-2905& 1413-2905   & 6926& 14 16 10.13& -29 19 23.8& 7.7 & 4.8& ND   & $<$ 1.9$^{B}$& P\\
14423-2039& N5734       & 4074& 14 45  8.93& -20 52 15.8& 25.1& 69.0& 108& $<$ 1.7$^{B}$& E\\
14423-2042& N5743       & 4216& 14 45 10.95& -20 54 46.7& 19.6& C    & C    & $<$ 2.0$^{B}$& E\\
14430-3728& 1443-3728   & 4476& 14 46 10.25& -37 41  5.9& 10.2& 15.5& 51& $<$ 1.7$^{A}$& P\\
14544-4255& I4518A M    & 4875& 14 57 41.21& -43 07 56.2& 43.0& 106.3& 151&   3.8$^{A}$& P\\
14566-1629& N5793       & 3521& 14 59 24.78& -16 41 37.0& 445.5& 833.3& 1129& 266.1$^{B}$& P\\
15226-4149& 1522-4149   & 4896& 15 25 58.39& -42 00 30.1& 4.3 & 13.7& 7& $<$ 1.8$^{A}$& P\\
15554-6614& 1555-6614 M & 3700& 16 00 01.59& -66 23 48.4& 1.4 & 20.6& 22& $<$ 1.7$^{A}$& P\\
15555-6610& I4585       & 3638& 16 00 16.3  & -66 19 16   & ND   & 9.9 & ND   & $<$ 1.5$^{A}$& P\\
16079-3111& 1607-3111   & 4451& 16 11  7.48& -31 19 13.8& 3.6 & 6.9 & ND   & $<$ 1.7$^{A}$& P\\
16153-7001& I4595       & 4532& 16 20 43.43& -70 08 35.5& 14.0& 42.0& 63& $<$ 1.3$^{A}$& P\\
16161-2419& 1616-2419   & 3542& 16 19  9.41& -24 26 47.5& 8.9 & 14.5& 29& $<$ 1.7$^{B}$& P\\
16229-6640& 1622-6640   & 6535& 16 27 51.67& -66 47 11.1& 9.8 & 4.6 & ND    & $<$ 1.4$^{A}$& P\\
16443-2916& 1644-2916   & 6264& 16 47 31.10& -29 21 22.0& 11.4& 19.7& 21&  NObs&  P\\
16516-0948& 1651-0948   & 6755& 16 54 23.98& -09 53 23.6& 20.7& ND   & ND   & $<$ 1.7$^{B}$& P\\
17138-1017& 1713-1017   & 5261& 17 16 35.87& -10 20 40.5& 25.8& 41.7& 59& $<$ 1.5$^{B}$& P\\
17182-7353& 1718-7353   & 4788& 17 24 36.64& -73 56 23.3& 12.2& 29.6& ND   & $<$ 1.7$^{A}$& P\\
17222-5953& 1722-5953   & 6241& 17 26 43.56& -59 55 56.7& 16.6& 30.3& 46& $<$ 1.7$^{A}$& P\\
17260-7622& 1726-7622   & 5507& 17 33 11.45& -76 24 46.4& 14.0& 19.4& 24& $<$ 1.6$^{A}$& P\\
18093-5744& I4687 M     & 5200& 18 13 39.65& -57 43 32.1& 31.2& 65.0& 100.& $<$ 1.4$^{A}$& E\\
18293-3413& 1829-3413   & 5449& 18 32 41.15& -34 11 27.5& 79.5& 121.1& 209& $<$ 2.1$^{A}$& E\\
18341-5732& I4734       & 4601& 18 38 25.69& -57 29 24.9& 23.8& 40.6& 62& $<$ 1.8$^{A}$& P\\
18421-5049& 1842-5049   & 5266& 18 46  2.68& -50 46 27.1& 9.7 & 11.6& 25& $<$ 1.5$^{A}$& P\\
18429-6312& I4769 M     & 4367& 18 47 43.90& -63 09 23.7& 7.9 & 18.5& 22& $<$ 1.8$^{A}$& P\\
19543-3804& 1954-3804   & 5713& 19 57 37.55& -37 56  8.9& 61.7& 76.4& 146&  23.4$^{A,B*}$& P\\
20305-0211& N6926       & 5970& 20 33 06.08& -02 01 37.1& 39.3& 65.1& 107& $<$ 1.7$^{B}$& E\\
20309-1132& N6931       & 3549& 20 33 41.43& -11 22 12.0& 8.8 & 13.8& ND   & $<$ 1.7$^{B}$& P\\
20486-4857& N6970       & 5288& 20 52  9.55& -48 46 39.9& 11.0& 30.1& 34& $<$ 1.7$^{A}$& E\\
21008-4347& 2100-4347   & 5208& 21 04 11.05& -43 35 35.3& 15.2& 27.4& 29& $<$ 1.7$^{A}$& P\\
21314-4102& N7087       & 5161& 21 34 33.41& -40 49  5.0& 8.1 & 15.5& 22& $<$ 1.4$^{A}$& P\\
21330-3846& 2133-3846AM & 5714& 21 36 10.74& -38 32 39.1& 10.2& 16.7& 21& $<$ 1.7$^{A}$& P\\
21453-3511& I5135       & 4842& 21 48 19.51& -34 57  4.3& 64.7& 122.4& 165&   4.7$^{A}$& E\\
22115-3013& 2211-3013   & 4291& 22 14 24.20& -29 58 53.3& 7.1 & 2.3& ND   & $<$ 1.7$^{A}$& P\\
22118-2742& 2211-2742   & 5247& 22 14 39.65& -27 27 50.8& 16.3& 24.6& 33.5& $<$ 2.6$^{B*}$& E\\
22179-2455& N7252       & 4688& 22 20 44.69& -24 40 43.4& 9.4 & ND   & ND   & $<$ 2.5$^{B*}$& P\\
23394-0353& 2339-0354   & 6966& 23 42 0.85  & -03 36 57.5& 17.4& ND   & ND   & $<$ 1.5$^{B}$& P\\
\enddata 
\end{deluxetable} 

\begin{deluxetable}{lccccccccc}
\tabletypesize{\scriptsize}
\tablewidth{0pc}
\tablecolumns{10}
\tablecaption{FIR and radio luminosities of the Southern COLA galaxies.}
\tablecomments {Luminosities were obtained from the fluxes quoted in Table 1 and given in units of WHz$^{-1}$ for ease of comparison. $q_{4.8}$,$q_{2.5}$ and $q_{1.4}$ are defined as the log of the ratio of the FIR flux to the radio flux at 4.8GHz, 2.5GHz and 1.4 GHz respectively and have associated errors of $\pm$ 0.05 at 4.8 and 2.5GHz and $\pm$ 0.07 at 1.3GHz.  The final column gives the spectral index, $\alpha$ between 4.8GHz and 1.4GHz, where flux at a given frequency, $\nu$, is defined as $S_{\nu}\propto \nu^{-\alpha}$. The estimated accuracy of $\alpha$ is $\pm$ 0.07.}
\tablehead{
\colhead{IRAS}  &     \colhead{Redshift}  &  \colhead{FIR}   &     \colhead{4.8GHz }   &  \colhead{2.5GHz}    &  \colhead{1.4 GHz }& \colhead{ q$_{4.8}$}  &  \colhead{q$_{2.5}$}  &  \colhead{q$_{1.4}$}  &  \colhead{$\alpha$}\\
\colhead{Name}    & \colhead{}     & \colhead{ (WHz$^{-1}$)}  &  \colhead{(WHz$^{-1}$)}      & \colhead{ (WHz$^{-1}$)}  &  \colhead{(WHz$^{-1}$)} &     &         &        &   \\}
\startdata
00085-1223  &  0.019  &  1.54E+25  &  2.15E+22  &  3.24E+22  &  4.24E+22  &  2.85  &  2.68  &  2.56  &  0.54\\
00344-3349  &  0.021  &  6.10E+24  &  1.24E+22  &  1.26E+22  &  \nodata   &  2.69  &  2.68  &  \nodata  &  \nodata\\
00402-2350  &  0.022  &  1.44E+25  &  2.17E+22  &  3.41E+22  &  5.96E+22  &  2.82  &  2.63  &  2.38  &  0.80\\
01053-1746  &  0.020  &  2.39E+25  &  5.80E+22  &  1.14E+23  &  1.64E+23  &  2.62  &  2.32  &  2.16  &  0.83\\
01159-4443  &  0.022  &  1.15E+25  &  2.08E+22  &  2.49E+22  &  3.91E+22  &  2.74  &  2.66  &  2.47  &  0.5\\
01165-1719  &  0.020  &  4.47E+24  &  8.65E+21  &  \nodata   &  2.26E+22  &  2.71  &  \nodata  &  2.30  &  0.76\\
01326-3623  &  0.016  &  5.65E+24  &  1.53E+22  &  2.61E+22  &  2.79E+22  &  2.57  &  2.33  &  2.31  &  0.47\\
01341-3734  &  0.017  &  6.18E+24  &  5.99E+21  &  1.45E+22  &  \nodata    &    3.01  &  2.63  &  \nodata  &  \nodata\\
01384-7515  &  0.013  &  3.68E+24  &  5.23E+21  &  8.44E+21  &  1.27E+22  &  2.85  &  2.64  &  2.46  &  0.71\\
02015-2333  &  0.016  &  3.16E+24  &  5.13E+21  &  1.24E+22  &  1.12E+22  &  2.79  &  2.41  &  2.45  &  0.62\\
02069-1022  &  0.014  &  3.25E+24  &  4.62E+21  &  \nodata   &  \nodata    &    2.85  &  \nodata  &  \nodata  &  \nodata\\
02072-1025  &  0.013  &  1.08E+25  &  6.16E+21  &  \nodata   &  \nodata    &    3.25  &  \nodata  &  \nodata  &  \nodata\\
02140-1134  &  0.013  &  3.21E+24  &  5.48E+21  &  1.25E+22  &  1.84E+22  &  2.77  &  2.41  &  2.24  &  0.96\\
02281-0309  &  0.019  &  7.36E+24  &  1.44E+22  &  1.43E+22  &  \nodata    &    2.71  &  2.71  &  \nodata  &  \nodata\\
02433-1534  &  0.014  &  4.00E+24  &  7.26E+21  &  1.29E+22  &  1.80E+22  &  2.74  &  2.49  &  2.35  &  0.72\\
02436-5556  &  0.018  &  5.23E+24  &  7.20E+21  &  1.10E+22  &  \nodata    &    2.86  &  2.68  &  \nodata  &  \nodata\\
02476-3858  &  0.017  &  3.32E+24  &  3.40E+21  &  \nodata   &  \nodata    &    2.99  &  \nodata  &  \nodata  &  \nodata\\
03022-1232  &  0.014  &  4.06E+24  &  5.50E+21  &  6.45E+21  &  4.61E+21  &  2.87  &  2.80  &  2.94  &  -0.14\\
04118-3207  &  0.012  &  5.20E+24  &  7.07E+21  &  1.11E+22  &  1.65E+22  &  2.87  &  2.67  &  2.50  &  0.67\\
04210-4042  &  0.020  &  9.65E+24  &  1.42E+22  &  2.64E+22  &  3.42E+22  &  2.83  &  2.56  &  2.45  &  0.70\\
04315-0840  &  0.016  &  1.86E+25  &  3.06E+22  &  4.16E+22  &  5.98E+22  &  2.78  &  2.65  &  2.49  &  0.53\\
04335-2514  &  0.016  &  3.97E+24  &  7.97E+21  &  1.14E+22  &  6.08E+21  &  2.70  &  2.54  &  2.81  &  -0.22\\
04370-2416  &  0.015  &  3.85E+24  &  6.57E+21  &  9.62E+21  &  2.17E+22  &  2.77  &  2.60  &  2.25  &  0.95\\
04461-0624  &  0.015  &  4.49E+24  &  1.06E+22  &  2.54E+22  &  2.47E+22  &  2.63  &  2.25  &  2.26  &  0.67\\
04502-3304  &  0.019  &  6.94E+24  &  6.14E+21  &  6.85E+21  &  7.46E+21  &  3.05  &  3.01  &  2.97  &  0.15\\
04558-0751  &  0.013  &  2.41E+24  &  4.08E+21  &  1.53E+22  &  2.50E+22  &  2.77  &  2.20  &  1.98  &  1.44\\
04569-0756  &  0.014  &  4.09E+24  &  5.49E+21  &  6.85E+21  &  1.14E+22  &  2.87  &  2.78  &  2.55  &  0.58\\
04591-0419  &  0.014  &  1.92E+24  &  3.42E+21  &  \nodata   &  \nodata   &    2.75  &  \nodata  &  \nodata  &  \nodata\\
04595-1813  &  0.013  &  2.21E+24  &  3.96E+21  &  1.86E+22  &  3.40E+22  &  2.75  &  2.07  &  1.81  &  1.71\\
05041-4938  &  0.014  &  2.46E+24  &  1.03E+21  &  1.47E+21  &  8.92E+21  &  3.38  &  3.22  &  2.44  &  1.72\\
05053-0805  &  0.015  &  5.42E+24  &  4.61E+21  &  6.46E+21  &  1.36E+22  &  3.07  &  2.92  &  2.60  &  0.86\\
05140-6213  &  0.017  &  3.65E+24  &  2.01E+21  &  5.99E+21  &  5.60E+21  &  3.26  &  2.78  &  2.81  &  0.81\\
05449-0651  &  0.022  &  7.61E+24  &  \nodata   &  \nodata   &  \nodata   &    \nodata  &  \nodata  &  \nodata  &  \nodata\\
05562-6933  &  0.015  &  2.68E+24  &  4.54E+21  &  8.50E+21  &  1.09E+22  &  2.77  &  2.50  &  2.39  &  0.70\\
06295-1735  &  0.021  &  8.59E+24  &  7.79E+21  &  1.57E+22  &  3.05E+22  &  3.04  &  2.74  &  2.45  &  1.09\\
06592-6313  &  0.023  &  7.76E+24  &  8.79E+21  &  1.22E+22  &  9.62E+21  &  2.95  &  2.80  &  2.91  &  0.07\\
08175-1433  &  0.019  &  5.10E+24  &  1.13E+22  &  2.28E+22  &  3.81E+22  &  2.66  &  2.35  &  2.13  &  0.97\\
08225-6936  &  0.013  &  2.09E+24  &  2.93E+21  &  3.65E+21  &  4.57E+21  &  2.85  &  2.76  &  2.66  &  0.35\\
08364-1430  &  0.014  &  2.36E+24  &  4.63E+21  &  7.31E+21  &  1.04E+22  &  2.71  &  2.51  &  2.36  &  0.64\\
08438-1510  &  0.018  &  3.58E+24  &  5.69E+21  &  9.58E+21  &  1.22E+22  &  2.80  &  2.57  &  2.47  &  0.61\\
09006-6404  &  0.022  &  5.21E+24  &  1.09E+22  &  1.55E+22  &  2.33E+22  &  2.68  &  2.53  &  2.35  &  0.61\\
09248-1918  &  0.016  &  3.19E+24  &  5.06E+21  &  7.96E+21  &  7.24E+21  &  2.80  &  2.60  &  2.64  &  0.29\\
09375-6951  &  0.020  &  5.97E+24  &  1.38E+22  &  2.11E+22  &  3.23E+22  &  2.64  &  2.45  &  2.27  &  0.68\\
10015-0614  &  0.017  &  9.98E+24  &  2.06E+22  &  4.14E+22  &  5.20E+22  &  2.68  &  2.38  &  2.28  &  0.73\\
10221-2317  &  0.012  &  4.11E+24  &  6.51E+21  &  1.10E+22  &  1.12E+22  &  2.80  &  2.57  &  2.56  &  0.43\\
10409-4557  &  0.023  &  9.55E+24  &  1.42E+22  &  \nodata   &  \nodata   &    2.83  &  \nodata  &  \nodata  &  \nodata\\
10484-0153  &  0.015  &  3.14E+24  &  \nodata   &  1.04E+22  &  \nodata   &    \nodata  &  2.48  &  \nodata  &  \nodata\\
10567-4310  &  0.017  &  5.04E+24  &  6.41E+21  &  1.12E+22  &  1.83E+22  &  2.90  &  2.65  &  2.44  &  0.83\\
11005-1601  &  0.013  &  3.52E+24  &  5.54E+21  &  1.45E+22  &  1.89E+22  &  2.80  &  2.38  &  2.27  &  0.98\\
11254-4120  &  0.016  &  5.10E+24  &  5.31E+21  &  7.39E+21  &  1.08E+22  &  2.98  &  2.84  &  2.67  &  0.57\\
11328-4844  &  0.019  &  4.34E+24  &  8.95E+21  &  1.45E+22  &  1.60E+22  &  2.69  &  2.47  &  2.43  &  0.46\\
11409-1631  &  0.012  &  1.73E+24  &  2.36E+21  &  5.11E+21  &  4.89E+21  &  2.86  &  2.53  &  2.55  &  0.58\\
12042-3140  &  0.023  &  1.05E+25  &  1.91E+22  &  3.01E+22  &  4.30E+22  &  2.74  &  2.54  &  2.39  &  0.65\\
12112-4659  &  0.018  &  4.16E+24  &  4.61E+21  &  5.94E+21  &  1.31E+22  &  2.96  &  2.85  &  2.50  &  0.83\\
12115-4657  &  0.018  &  9.65E+24  &  1.19E+22  &  7.97E+21  &  2.50E+22  &  2.91  &  3.08  &  2.59  &  0.59\\
12120-1118  &  0.018  &  3.76E+24  &  7.35E+21  &  1.45E+22  &  1.88E+22  &  2.71  &  2.41  &  2.30  &  0.75\\
12171-1156  &  0.014  &  2.34E+24  &  1.96E+21  &  4.92E+21  &  5.78E+21  &  3.08  &  2.68  &  2.61  &  0.86\\
12286-2600  &  0.020  &  4.58E+24  &  5.07E+21  &  \nodata   &  \nodata   &    2.96  &  \nodata  &  \nodata  &  \nodata\\
12329-3938  &  0.012  &  1.79E+24  &  4.70E+21  &  1.03E+22  &  1.52E+22  &  2.58  &  2.24  &  2.07  &  0.93\\
12351-4015  &  0.018  &  6.89E+24  &  9.46E+21  &  1.75E+22  &  2.89E+22  &  2.86  &  2.59  &  2.38  &  0.89\\
12504-2711  &  0.012  &  1.97E+24  &  3.25E+21  &  \nodata   &  1.29E+22  &  2.78  &  \nodata  &  2.18  &  1.09\\
12596-1529  &  0.016  &  4.70E+24  &  6.04E+21  &  1.18E+22  &  1.39E+22  &  2.89  &  2.60  &  2.53  &  0.66\\
13001-2339  &  0.022  &  1.59E+25  &  3.28E+22  &  4.47E+22  &  7.62E+22  &  2.69  &  2.55  &  2.32  &  0.67\\
13035-4008  &  0.015  &  3.34E+24  &  5.42E+21  &  8.25E+21  &  1.04E+22  &  2.79  &  2.61  &  2.51  &  0.52\\
13097-1531  &  0.021  &  1.43E+25  &  2.58E+22  &  4.54E+22  &  7.71E+22  &  2.74  &  2.50  &  2.27  &  0.87\\
13135-2801  &  0.015  &  3.14E+24  &  3.70E+21  &  7.02E+21  &  8.73E+21  &  2.93  &  2.65  &  2.56  &  0.68\\
13193-5208  &  0.017  &  3.57E+24  &  2.81E+21  &  4.29E+21  &  \nodata   &    3.10  &  2.92  &  \nodata  &  \nodata\\
13197-1627  &  0.017  &  3.96E+24  &  5.57E+22  &  1.03E+23  &  1.47E+23  &  1.85  &  1.58  &  1.43  &  0.77\\
13229-2934  &  0.014  &  8.76E+24  &  2.50E+22  &  4.33E+22  &  7.14E+22  &  2.54  &  2.31  &  2.09  &  0.83\\
13301-2357  &  0.016  &  4.51E+24  &  6.79E+21  &  1.64E+22  &  3.15E+22  &  2.82  &  2.44  &  2.16  &  1.22\\
13322-3500  &  0.013  &  2.85E+24  &  5.14E+21  &  9.52E+21  &  1.17E+22  &  2.74  &  2.48  &  2.39  &  0.65\\
14132-2905  &  0.023  &  5.77E+24  &  7.91E+21  &  4.93E+21  &  \nodata   &    2.86  &  3.07  &  \nodata  &  \nodata\\
14423-2039  &  0.014  &  6.62E+24  &  8.91E+21  &  2.45E+22  &  3.83E+22  &  2.87  &  2.43  &  2.24  &  1.16\\
14423-2042  &  0.014  &  7.11E+24  &  7.46E+21  &  \nodata   &  \nodata   &    2.98  &  \nodata  &  \nodata  &  \nodata\\
14430-3728  &  0.015  &  3.39E+24  &  4.38E+21  &  6.65E+21  &  2.19E+22  &  2.89  &  2.71  &  2.19  &  1.28\\
14544-4255  &  0.016  &  6.00E+24  &  2.19E+22  &  5.42E+22  &  7.68E+22  &  2.44  &  2.04  &  1.89  &  1.00\\
14566-1629  &  0.012  &  2.35E+24  &  1.18E+23  &  2.21E+23  &  2.99E+23  &  1.30  &  1.03  &  0.90  &  0.74\\
15226-4149  &  0.016  &  3.28E+24  &  2.21E+21  &  7.03E+21  &  3.75E+21  &  3.17  &  2.67  &  \nodata  &  \nodata\\
15554-6614  &  0.012  &  3.14E+24  &  3.98E+20  &  6.04E+21  &  6.38E+21  &  3.90  &  2.72  &  2.69  &  2.20\\
15555-6610  &  0.012  &  2.93E+24  &  \nodata   &  2.80E+21  &  \nodata   &    \nodata  &  3.02  &  \nodata  &  \nodata\\
16079-3111  &  0.015  &  2.58E+24  &  1.52E+21  &  2.92E+21  &  \nodata   &    3.23  &  2.95  &  \nodata  &  \nodata\\
16153-7001  &  0.012  &  3.21E+24  &  3.75E+21  &  1.13E+22  &  1.69E+22  &  2.93  &  2.45  &  2.28  &  1.20\\
16161-2419  &  0.015  &  4.21E+24  &  3.90E+21  &  6.38E+21  &  1.28E+22  &  3.03  &  2.82  &  2.52  &  0.94\\
16229-6640  &  0.022  &  6.70E+24  &  9.01E+21  &  4.22E+21  &  \nodata   &    2.87  &  3.20  &  \nodata  &  \nodata\\
16443-2916  &  0.021  &  1.05E+25  &  9.60E+21  &  1.66E+22  &  1.75E+22  &  3.04  &  2.80  &  2.78  &  0.48\\
16516-0948  &  0.023  &  1.04E+25  &  2.03E+22  &  \nodata   &  \nodata   &    2.71  &  \nodata  &  \nodata  &  \nodata\\
17138-1017  &  0.018  &  1.19E+25  &  1.51E+22  &  2.48E+22  &  3.52E+22  &  2.90  &  2.68  &  2.53  &  0.67\\
17182-7353  &  0.016  &  3.78E+24  &  6.01E+21  &  1.45E+22  &  \nodata   &    2.80  &  2.42  &  \nodata  &  \nodata\\
17222-5953  &  0.021  &  1.00E+25  &  1.39E+22  &  2.54E+22  &  3.87E+22  &  2.86  &  2.60  &  2.41  &  0.81\\
17260-7622  &  0.018  &  2.57E+24  &  9.11E+21  &  1.26E+22  &  1.56E+22  &  2.45  &  2.31  &  2.22  &  0.43\\
18093-5744  &  0.017  &  1.55E+25  &  1.81E+22  &  3.77E+22  &  5.80E+22  &  2.93  &  2.61  &  2.43  &  0.93\\
18293-3413  &  0.018  &  3.22E+25  &  5.06E+22  &  7.71E+22  &  1.33E+23  &  2.80  &  2.62  &  2.38  &  0.77\\
18341-5732  &  0.015  &  9.69E+24  &  1.08E+22  &  1.84E+22  &  2.81E+22  &  2.95  &  2.72  &  2.54  &  0.76\\
18421-5049  &  0.018  &  4.35E+24  &  5.77E+21  &  6.89E+21  &  1.49E+22  &  2.88  &  2.80  &  2.47  &  0.75\\
18429-6312  &  0.015  &  2.78E+24  &  3.24E+21  &  7.55E+21  &  9.09E+21  &  2.93  &  2.57  &  2.48  &  0.82\\
19543-3804  &  0.019  &  5.64E+24  &  4.32E+22  &  5.35E+22  &  1.02E+23  &  2.12  &  2.02  &  1.74  &  0.68  \\
20305-0211  &  0.020  &  7.68E+24  &  3.01E+22  &  4.98E+22  &  8.16E+22  &  2.41  &  2.19  &  1.97  &  0.79  \\
20309-1132  &  0.012  &  1.72E+24  &  2.38E+21  &  3.71E+21  &  \nodata   &    2.86  &  2.67  &  \nodata  &  \nodata  \\
20486-4857  &  0.018  &  4.23E+24  &  6.61E+21  &  1.80E+22  &  2.05E+22  &  2.81  &  2.37  &  2.31  &  0.90 \\
21008-4347  &  0.017  &  7.01E+24  &  8.82E+21  &  1.60E+22  &  1.71E+22  &  2.90  &  2.64  &  2.61  &  0.52  \\
21314-4102  &  0.017  &  3.89E+24  &  4.61E+21  &  8.88E+21  &  1.27E+22  &  2.93  &  2.64  &  2.49  &  0.80  \\
21330-3846  &  0.019  &  6.08E+24  &  7.16E+21  &  1.17E+22  &  1.44E+22  &  2.93  &  2.72  &  2.63  &  0.56  \\
21453-3511  &  0.016  &  1.19E+25  &  3.25E+22  &  6.15E+22  &  8.30E+22  &  2.56  &  2.29  &  2.16  &  0.75  \\
22115-3013  &  0.014  &  2.04E+24  &  2.80E+21  &  9.19E+20  &  \nodata   &    2.86  &  3.35  &  \nodata  &  \nodata  \\
22118-2742  &  0.018  &  5.32E+24  &  9.64E+21  &  1.45E+22  &  1.98E+22  &  2.74  &  2.56  &  2.43  &  0.57  \\
22179-2455  &  0.016  &  2.86E+24  &  4.44E+21  &  \nodata   &  \nodata   &    2.81  &  \nodata  &  \nodata  &  \nodata  \\
23394-0353  &  0.023  &  9.96E+24  &  1.82E+22  &  \nodata   &  \nodata   &    2.74  &  \nodata  &  \nodata  &  \nodata  \\
\enddata 
\end{deluxetable} 

\begin{deluxetable}{lcccccc}
\tabletypesize{\footnotesize}
\tablewidth{0pc}
\tablecolumns{7}
\tablecaption{The median values of q, $\langle$q$\rangle$, for the different samples of galaxies.}
\tablecomments{The standard deviation ($\sigma$) of q is given in brackets next to the median value $\langle$q$\rangle$. ``LBA non-detections'' indicates sources observed with the LBA but not detected, ``LBA all detections'' refers to the 9 sources in which compact cores were detected. IRAS 12329-3938 is excluded from the ``LBA strong detections'' sample. }
\tablehead{
\colhead{Sample}  &\colhead{N$_{4.8}$}&\colhead{$\langle$q$_{4.8}\rangle$ ($\sigma$)}&\colhead{N$_{2.5}$}&\colhead{$\langle$q$_{2.5}\rangle$ ($\sigma$)}&\colhead{N$_{1.4}$}&\colhead{$\langle$q$_{1.4}\rangle$ ($\sigma$)}\\}
\startdata
All sources           &  104 &  2.83 (0.27) &  94 &  2.60 (0.31) &  84 &  2.43 (0.31)\\
LBA non-detection     &  93  &  2.85 (0.19) &  83 &  2.61 (0.23) &  73 &  2.44 (0.22)\\
LBA all detections    &  9   &  2.56 (0.51) &  9  &  2.24 (0.51) &  9  &  2.07 (0.51)\\
LBA strong detections &  8   &  2.44 (0.55) &  8  &  2.17 (0.54) &  8  &  2.03 (0.54)\\
\enddata 
\end{deluxetable}

\begin{deluxetable}{lccccccccccl}
\tabletypesize{\scriptsize}
\tablewidth{0pc}
\tablecolumns{12}
\tablecaption{Properties of the galaxies in which compact radio cores were detected.}
\tablecomments{$\alpha$ is the spectral index of the total radio luminosity of the galaxy (as measured on the ATCA) between 4.8GHz and 1.4GHz. R is the ratio of the excess radio emission to that predicted from the radio-FIR correlation (see discussion in text for further details). The spectral classes are defined as follows: Sy1=seyfert 1 galaxy; Sy2= Seyfert 2 galaxy; HII= starburst galaxy; B= galaxy lies close (less than 1 dex) to the division between AGN and starburst galaxies. References for the classification are (1) Corbett et al., in preparation, (2) Durret \& Bregeron (1986), (3) Baan, Saltzer \& Le Winter (1998), (4) Aguerro et al. (1994) (5) Heisler, Lumsden \& Bailey (1997) }
\tablehead{
\colhead{IRAS}& \colhead{LBA Flux} & \colhead{Error} & \colhead{L$_{core}$} & \colhead{Log($\frac{L_{core}}{L\sun}$)} & \colhead{Log(T$_{B}$)} &\colhead{q$_{4.8}$}& \colhead{q$_{2.5}$}&\colhead{q$_{1.4}$}&\colhead{$\alpha$}& \colhead{R}& \colhead{Spectral} \\   
\colhead{Name}&\colhead{(mJy)}  & \colhead{(mJy)} & \colhead{(W Hz$^{-1}$)}& \colhead{} & \colhead{} &  \colhead{} & \colhead{ } & \colhead{}&  \colhead{ } &  \colhead{ }& \colhead{Class} \\}
\startdata
09375-6951  &  3.6  &  0.21  &  28$\times10^{20}$  &  4.25  &  5.67  &  2.64  &  2.45  &  2.27  &  0.68  &  0.63  &  B$^{1}$\\
12329-3938  &  1.6  &  0.17  &  4.3$\times10^{20}$  &  3.43  &  5.33  &  2.58  &  2.24  &  2.07  &  0.93  &  0.85  &  Sy2$^{1,2}$\\
13097-1531  &  2.5  &  0.18  &  22$\times10^{20}$  &  4.14  &  5.52  &  2.74  &  2.50  &  2.27  &  0.87  &  0.27  &  B$^{1,3}$\\
13135-2801  &  2.2  &  0.15  &  9.0$\times10^{20}$  &  3.75  &  5.45  &  2.93  &  2.65  &  2.56  &  0.68  &  -0.17  &  HII$^{1}$\\
13197-1627  &  8.9  &  0.22  &  51$\times10^{20}$  &  4.50  &  6.07  &  1.85  &  1.58  &  1.43  &  0.77  &  8.9  &  Sy1.8$^{1,4}$\\
14544-4255  &  3.8  &  0.21  &  19$\times10^{20}$  &  4.08  &  5.69  &  2.44  &  2.04  &  1.89  &  1.00  &  1.6  &  Sy2$^{1}$\\
14566-1629  &  266  &  2.20  &  710$\times10^{20}$  &  5.64  &  7.54  &  1.30  &  1.03  &  0.90  & 0.74  &  34  &  Sy2$^{1,3}$\\
19543-3804  &  23.4 &  0.24  &  160$\times10^{20}$  &  5.01  &  6.49  &  2.11  &  2.02  &  1.74  &  0.68  &  4.4  &  Sy2$^{1}$\\
21453-3511  &  4.7  &  0.20  &  22$\times10^{20}$  &  4.17  &  5.79  &  2.56  &  2.29  &  2.16  &  0.75  &  0.93  &  Sy2$^{1,5}$\\
\enddata 
\end{deluxetable}

\begin{figure}
\figurenum{1}
\epsscale{0.8}
\plotone{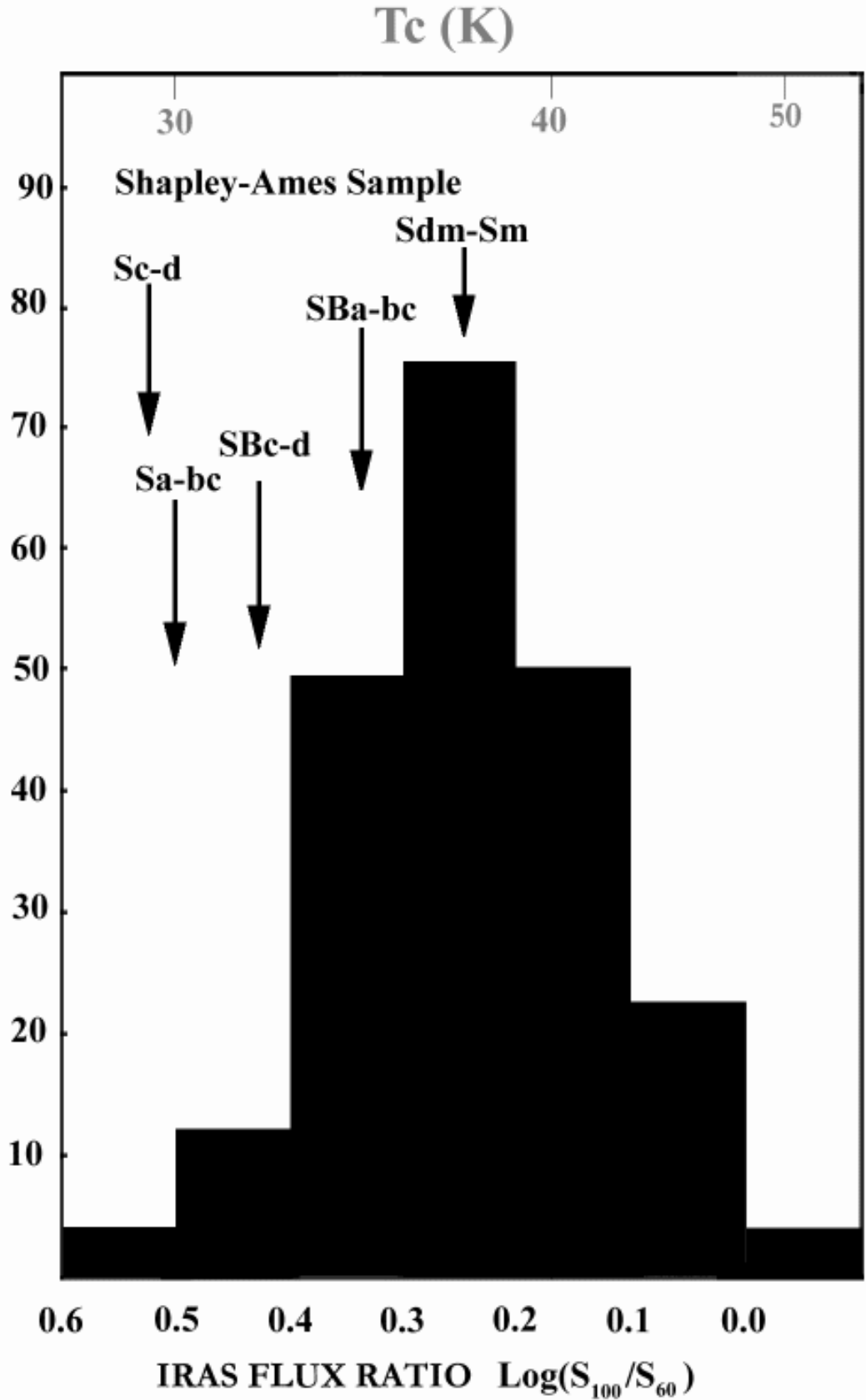}
\caption[]{The color-temperature distribution for the whole COLA sample (217 galaxies). The arrows show the mean values for the same ratio from the Shapley - Ames sample of de Jong et al. (1984) which is a non-interacting nearby
galaxy sample based on the first IRAS results. It can be seen that the peak
in the COLA distribution is much warmer than the de Jong et al. sample, which peaks at log(S$_{100}$/S$_{60}$) =0.5.}
\end{figure}
\begin{figure}
\rotate
\figurenum{2}
\epsscale{0.6}
\plotone{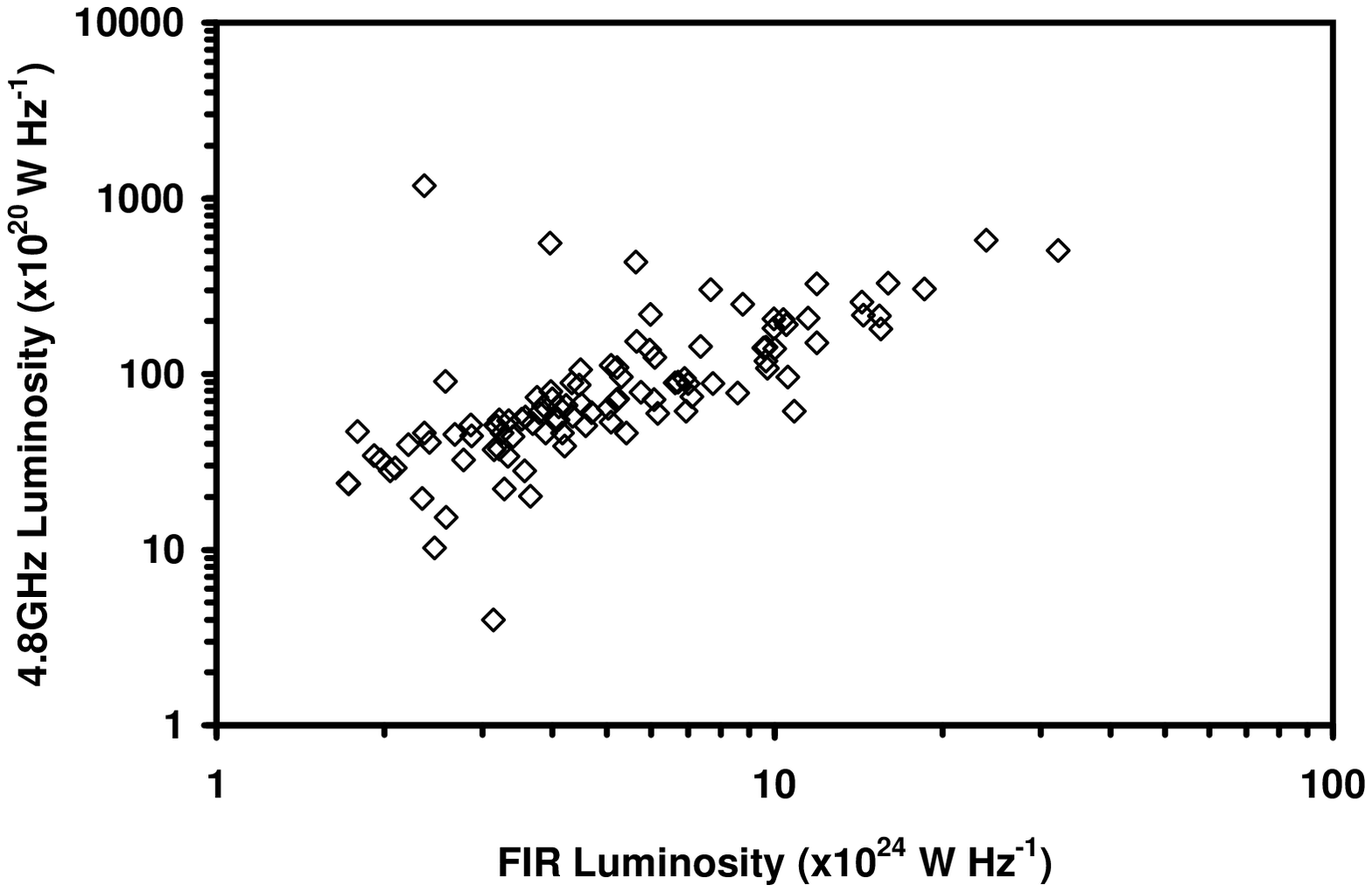}
\plotone{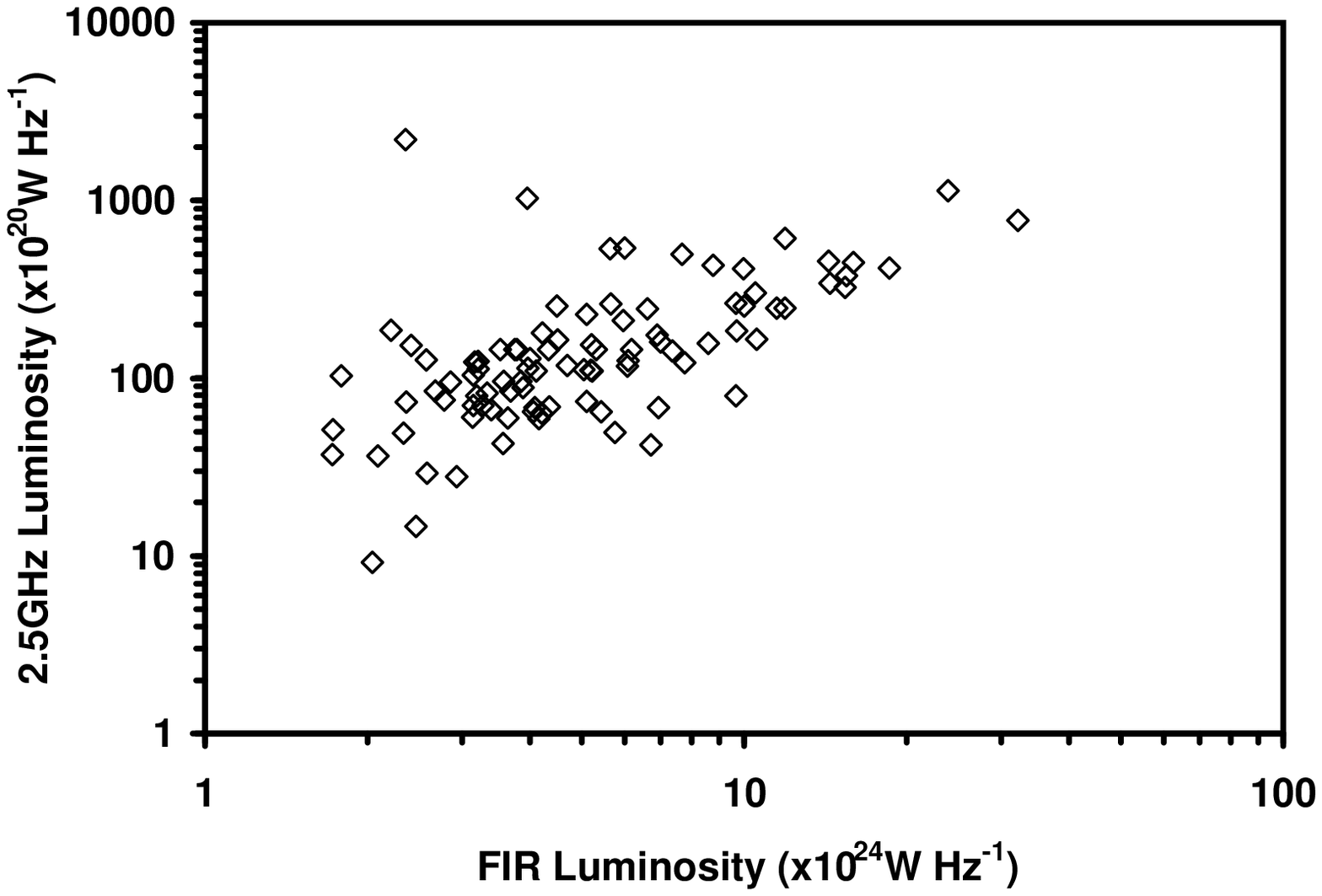}
\plotone{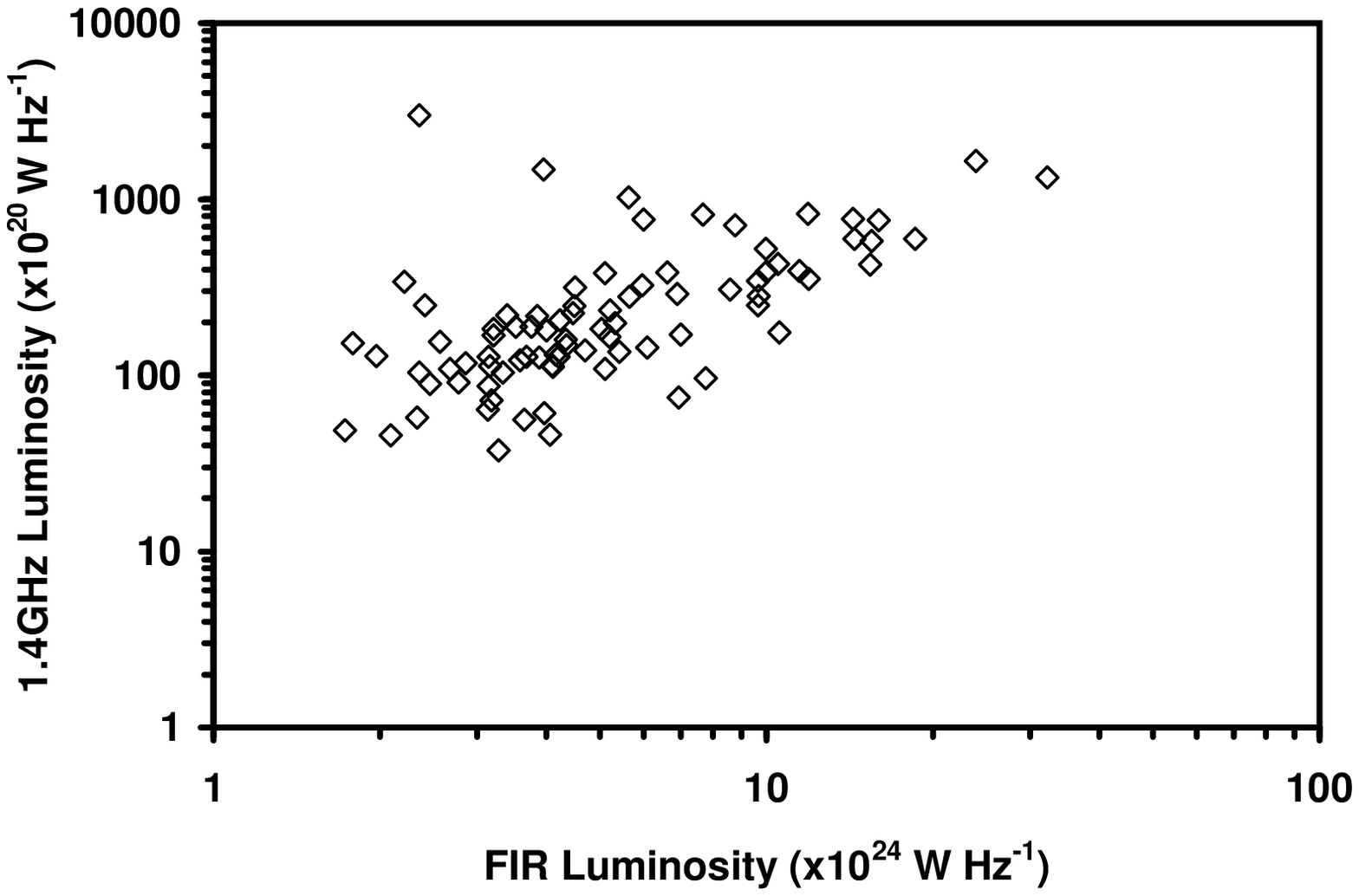}
\caption[]{Radio-FIR correlation of the COLA galaxies}
\end{figure}
\begin{figure}
\figurenum{3}
\epsscale{0.7}
\plotone{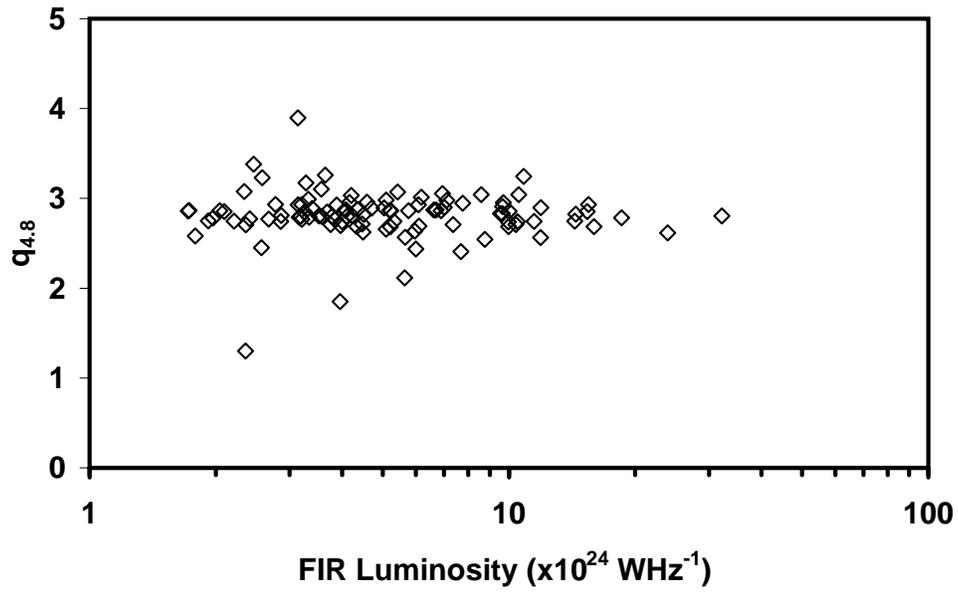}
\caption[]{q$_{4.8}$ versus FIR luminosity. No evidence is seen for a dependence of q$_{4.8}$ on the FIR luminosity.}
\end{figure}
\begin{figure}
\figurenum{4}
\epsscale{0.7}
\plotone{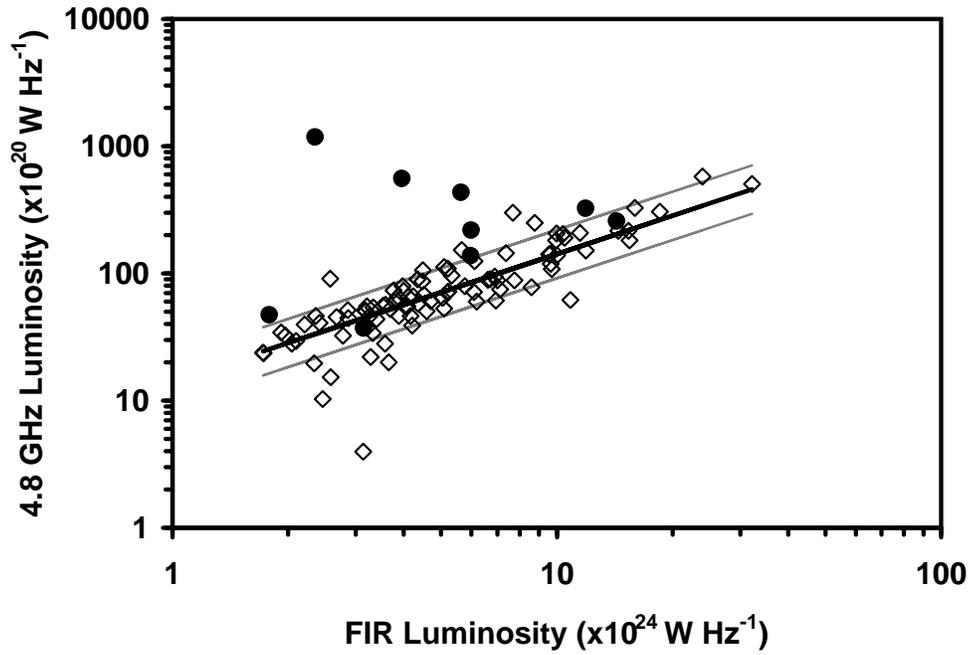}
\caption[]{Radio-FIR correlation at 4.8GHz. Filled circles indicate the galaxies with compact cores, open diamonds indicate the galaxies in which no LBA cores were detected. Solid lines are shown at $\langle q_{4.8} \rangle$ and $\langle q_{4.8} \rangle\pm \sigma$ where $\sigma$ is the standard deviation of $q_{4.8}$}
\end{figure}
\begin{figure}
\figurenum{5}
\epsscale{0.7}
\plotone{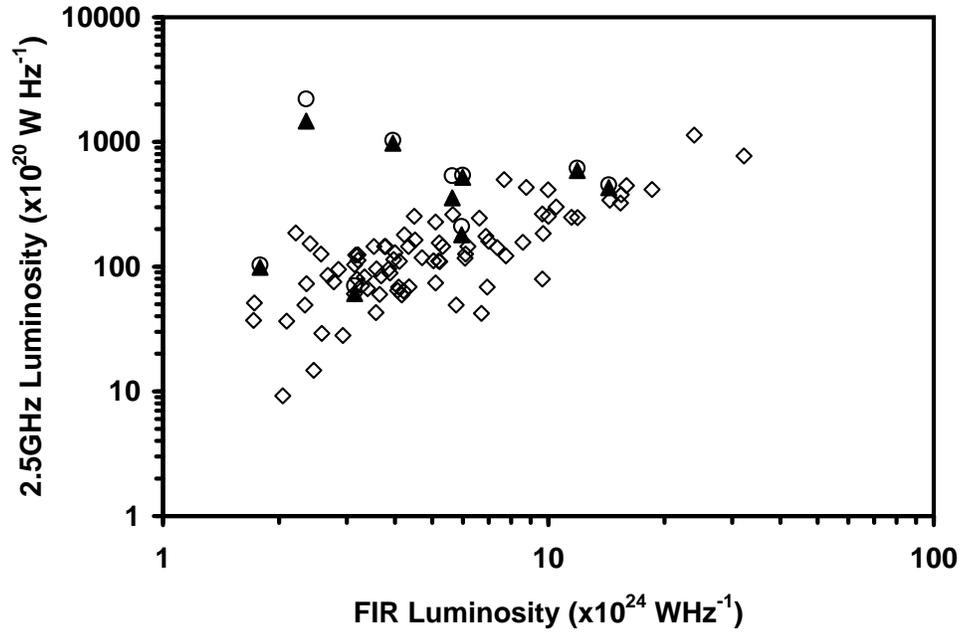}
\caption[]{Radio-FIR correlation at 2.5GHz. Open circles indicate the galaxies in which cores were detected and diamonds indicate galaxies in which LBA cores were not detected. Filled triangles indicate where the galaxies with compact cores move to on the radio-FIR correlation after subtraction of the core flux. Solid lines are shown at $\langle q_{2.5} \rangle$ and $\langle q_{2.5} \rangle\pm \sigma$ where $\sigma$ is the standard deviation of $q_{2.5}$}
\end{figure}
\begin{figure}
\rotate
\figurenum{6}
\epsscale{0.8}
\plotone{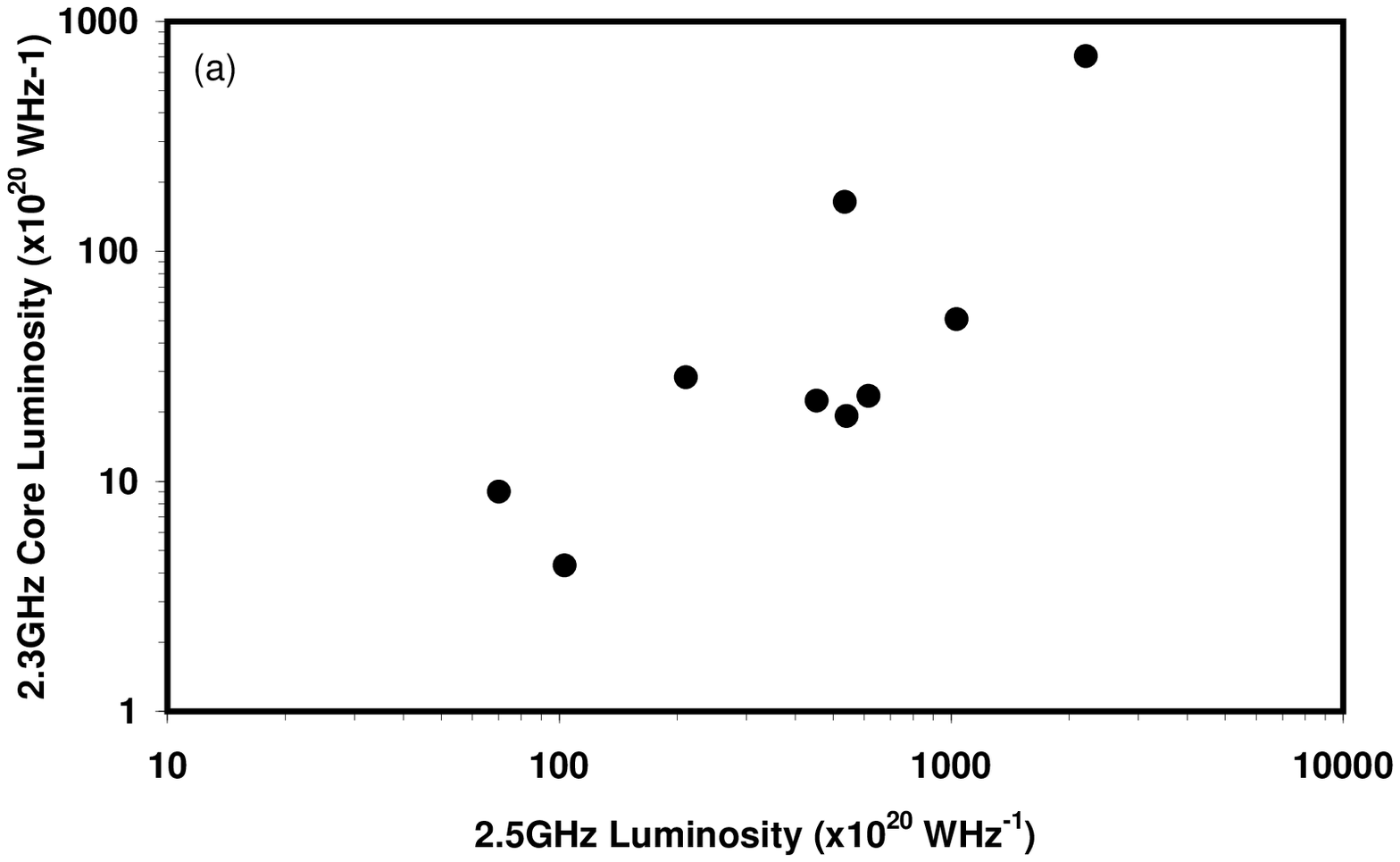}
\epsscale{0.8}
\plotone{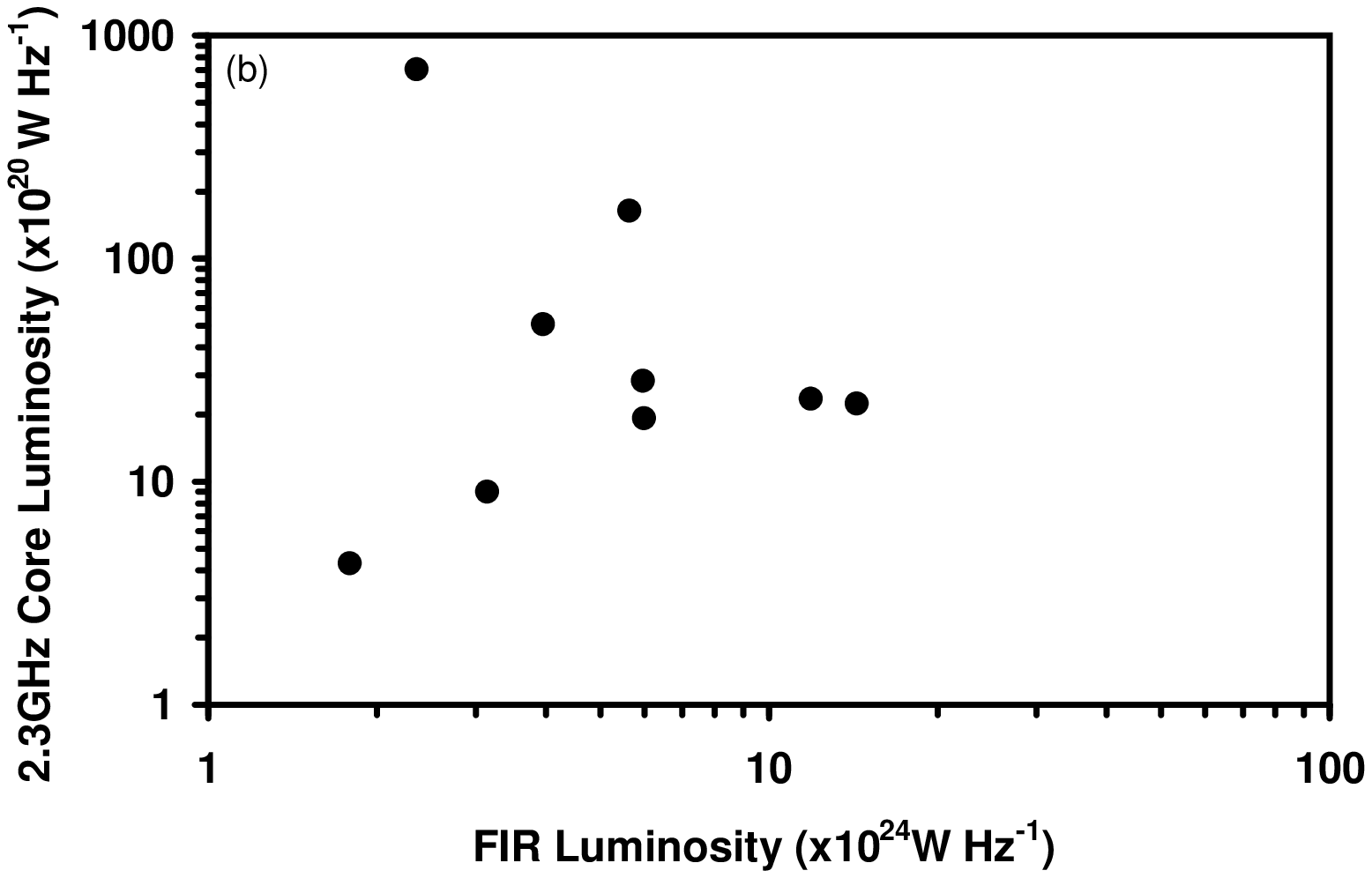}
\caption[]{Correlation of the 2.3GHz luminosity of the compact core with the total radio luminosity of the galaxy (a) and the FIR luminosity (b). A significant correlation can be seen with the total radio luminosity of the galaxy but there doesn not appear to be a correlation between the compact core luminosity and the FIR luminosity of the galaxy.}

\end{figure}

\end{document}